\documentstyle[12pt,rotate]{article}

\input epsf.sty

\newcommand{\inclfig}[2]{\mbox{\epsfxsize=#1cm \epsfbox{#2.ps}}}
\newcommand{\insertfig}[2]{\mbox{\epsfxsize=#1cm \epsfbox{#2.eps}}}
\newcommand{\res}{\mathop{\rm Res}}

\newcommand{\g}{{\sl g}}

\newcommand{\cD}{{\cal D}}

\newcommand{\cG}{{\cal G}}

\newcommand{\cV}{{\cal V}}

\newcommand{\cO}{{\cal O}}
\newcommand{\cH}{{\cal H}}

\newcommand{\cP}{{\cal P}}

\newcommand{\cQ}{{\cal Q}}
\newcommand{\cF}{{\cal F}}
\newcommand{\cE}{{\cal E}}
\newcommand{\cS}{{\cal S}}
\newcommand{\cT}{{\cal T}}
\newcommand{\cY}{{\cal Y}}
\newcommand{\OO}{\mathop{\otimes}}
\font\cmss=cmss12 
\def\1{\hbox{{1}\kern-.25em\hbox{l}}}
\def\bfZ{\relax{\hbox{\cmss Z\kern-.4em Z}}}

\begin{document}
\begin{titlepage}

\centerline{\large \bf Renormalization of twist-three operators}
\centerline{\large \bf and integrable lattice models.}

\vspace{10mm}

\centerline{\bf A.V. Belitsky\footnote{Alexander von Humboldt Fellow.}}

\vspace{10mm}

\centerline{\it Institut f\"ur Theoretische Physik, Universit\"at
                Regensburg}
\centerline{\it D-93040 Regensburg, Germany}

\vspace{20mm}

\centerline{\bf Abstract}

\vspace{0.8cm}

We address the problem of solution of the QCD three-particle evolution
equations which govern the $Q^2$-dependence of the chiral-even
quark-gluon-quark and three-gluon correlators contributing to a number
of asymmetries at leading order and the transversely polarized structure
function $g_2 (x_{\rm Bj})$. The quark-gluon-quark case is completely
integrable in multicolour limit and corresponds to a spin chain with
non-periodic boundary conditions, while the pure gluonic sector contains,
apart from a piece in the Hamiltonian equivalent to XXX Heisenberg magnet
of spin $s = - \frac{3}{2}$, a non-integrable addendum which can be
treated perturbatively for a bulk of the spectrum except for a few lowest 
energy levels. We construct a quasiclassical expansion with respect to the 
total conformal spin of the system and describe fairly well the energy 
spectra of quark-gluon-quark and three-gluon systems.

\vspace{3.5cm}

\noindent Keywords: twist-three operators, evolution, three-particle
problem, integrability, spectrum of eigenvalues

\vspace{0.5cm}

\noindent PACS numbers: 11.10.Gh, 11.15.Kc, 11.30.Na, 12.38.Cy

\end{titlepage}

\section{Twist-three effects in high energy scattering.}

The naive parton model \cite{Fey71} provided a first successful
microscopic description of several high energy inclusive processes
and endowed them with intuitive probabilistic interpretation as the
scattering of a probe on an incoherent bunch of free and collinear
constituents (quark and gluons) in a hadron target by expressing the
cross sections in terms of the densities of partons in a (incoming)
hadron or hadrons in a (outgoing) parton. Later it was realized that
this is only the first asymptotic term in the infinite series in
inverse powers of a hard momentum scale $Q$ where all quantum
mechanical interference effects enter the game as power corrections
and, loosely speaking, can be interpreted as an overlap of hadron wave
functions with different number of partons. For the deep inelastic
scattering this series has a firm field theoretical ground in the form
of the operator product expansion (OPE) \cite{Wil69,Col84}. For other
reactions which do not admit the OPE a set of factorization theorems
\cite{ColSopSte89} has been proven within the framework of QCD, however,
not far beyond the first non-leading term \cite{QiuSte91a} which goes
under the name of higher twist. These are the latter terms which are of
particular interest since being correlations of more than two field
operators they can give new insights into the QCD dynamics in the
non-perturbative domain which is still out of a systematic theoretical
control.

Typical higher twists in hard reactions are usually associated with
power suppressed contributions in a large momentum scale on the
background of a dominating twist-two cross section and while being
important cannot be studied experimentally with a high accuracy. However,
it is not always the case as there exists a class of transverse spin
dependent processes where the relevant observable is an asymmetry,
$\Delta \sigma = \sigma (\vec{s}_\perp) - \sigma (- \vec{s}_\perp)$,
(not a cross section $\sigma = \sigma (\vec{s}_\perp) +
\sigma (- \vec{s}_\perp)$) and where the twist-three contributes
as a leading effect not contaminated by a lower twist. The well-known
twist-three structure function $g_2 (x_{\rm Bj})$ \cite{Jaf92} measured
in the polarized deep inelastic scattering \cite{E143} is their good
representative. It acquires an operator definition via a Fourier transform
of the matrix element of a non-local operator of the quark fields 
separated by a light-like distance\footnote{We have omitted here (and 
everywhere below) the path ordered exponential assuming the light-cone 
gauge $B_+ = 0$.}
\begin{equation}
g_T (x) \equiv g_1 (x) + g_2 (x)
= \frac{1}{2} \int \frac{d \kappa}{2 \pi} e^{i \kappa x}
\langle h |
\bar\psi (0) \gamma^\perp \gamma_5 \psi (\kappa n)
| h \rangle .
\end{equation}
Being of the form similar to the conventional unpolarized distributions
this quantity is implicitly interaction dependent. This can be made
manifest by virtue of QCD equations of motion and the Lorentz invariance. 
Its genuine twist-three part can be reduced in leading order in the QCD
coupling constant to an integral of a quark-gluon-quark correlation function
(whose precise definition is given in Eq.\ (\ref{Ydefintion})) depending
on two momentum fractions $Y (x, x') \sim \langle \bar\psi G \psi \rangle$,
namely \cite{BukKurLip84}
\begin{equation}
\label{gTtoY}
g_2^{\rm tw-3} (x)
=
\int_{x}^{1} \frac{d x'}{x'}
\int
\frac{d x''}{x'' - x'}
\left[
\frac{\partial}{\partial x'} Y (x', x'')
+\frac{\partial}{\partial x''} Y (x'', x')
\right] .
\end{equation}
Evaluating the correction in $\alpha_s$ to the structure function $g_2$
requires the introduction of three-gluon operators $G \widetilde{G} G$
(see Eq.\ (\ref{PositionSpace})) which enter through the quark loop
coupling\footnote{Note, that the results of this reference are not
complete since the contribution from twist-three two-gluon operators
which are related to the three-gluon ones \cite{KodNasTocTan98} were
not taken into account and must be included for complete treatment.}
\cite{BelEfrTer95} and might be responsible for the small-$x_{\rm Bj}$
behaviour of $g_2 (x_{\rm Bj})$, --- the issue which is under debate
\cite{IvaNikProSch99}.

Another set of observables is single transverse spin asymmetries in
hadronic reactions where only one particle in the initial state is
polarized (or only polarization of a single particle in the final
state is tagged), i.e.\ the inclusive Compton scattering $\gamma \vec{p}
\to \gamma X$ \cite{EfrTer85}, the direct photon \cite{QiuSte91b,Ji92},
pion \cite{QiuSte91b} or jet production in $p\vec{p}$-collisions, the
Drell-Yan process with measured azimuthal angular dependence of a lepton
\cite{BoeMulTer89} etc. They are of great interest since they involve
almost the same quark-gluon-quark and three-gluon correlation functions
discussed above but not folded with a coefficient function and thus
depend on both (although equal) momentum fractions. This can be used
for their extraction or at least for a direct confrontation with
phenomenological models available.

Since the data are normally can not be taken at the same values of the
hard momentum variable $Q$, the correlation functions which enter cross 
sections are measured with different resolutions and, thus, they differ 
for different scales. Therefore, the question naturally appears about 
their relation. Their dependence on $Q$ is only logarithmic and within 
the context of QCD it results from the behaviour of the theory at the 
cut-off. This can be consistently described within the framework of the 
renormalization group for composite operators. In the language of the QCD 
improved parton model the task of its description is translated into the 
construction and solution of the evolution equation for the multi-parton
correlation functions. At leading order in the coupling constant, to which
we restrict ourselves, the former problem is relatively simple and can be
resolved in a straightforward manner \cite{BukFroLipKur85,Bel97,KodTan98}.
However, the second issue is by far more complicated since we face here a
three-particle problem and a priori there is a little hope to find
analytical solutions to it. The ultimate result of the scale dependence
of a correlator is expected to be of the form
\begin{equation}
F ( x, x' | Q^2 )
= \sum_{\{ \alpha \}}
{\mit\Psi}_{\{ \alpha \}} (x, x')
\left(
\frac{\alpha_s ( Q_0^2 )}{\alpha_s ( Q^2 )}
\right)^{f_{(c)} \cE_{\{ \alpha \}} / \beta_0}
\langle\langle \cF_{\{ \alpha \}} ( Q_0^2 ) \rangle\rangle ,
\end{equation}
where $\cE_{\{\alpha\}}$ are the eigenvalues of evolution kernels (and
$f_{(c)}$ is an extracted colour factor) and ${\mit\Psi}_{\{ \alpha \}}
(x, x')$ are the corresponding eigenfunctions parametrized by a set
quantum numbers $\{ \alpha \}$. As usual $\beta_0 = \frac{4}{3} T_F N_f
- \frac{11}{3} C_A$ is the leading term of the QCD $\beta$-function
and $\langle\langle \cF_{\{ \alpha \}} ( Q_0^2 ) \rangle\rangle$ stand
for reduced matrix elements of local operators at a low normalization
point $Q_0$.

Very recently a guiding principle has been found which has allowed to
cope the second problem and it is based on the integrability of
spin lattice models to which the corresponding evolution equations
can be reduced \cite{BraDerMan98,BraDerKorMan99,Bel99a,Bel99b}. These
findings are analogous to a similar equivalence encountered in the
problem of description of the Regge behaviour of QCD amplitudes
\cite{Lip94,FadKor95,Kor95a,Kor95b}. The studies have been undertaken
for the three-quark \cite{BraDerMan98,BraDerKorMan99} and chiral-odd
quark-gluon-quark \cite{BraDerMan98,Bel99a,Bel99b} operators and allowed
to find the anomalous dimensions, hereafter referred to as energies, of
the eigenstates as well as their eigenfunctions in a WKB type manner
where the r\^ole of the Planck constant has been played by the inverse
total conformal spin of the system. These results compared quite
well with an explicit numerical diagonalization of the mixing matrix
and, therefore, they provided a reasonable approximation.

In the present paper we continue the study along this line and
give almost complete description of the three-gluon and
quark-gluon-quark chiral even sector. Our presentation will
be organized as follows. Next section is devoted to the study
of the purely gluonic evolution equation. This means that we discard
completely from our consideration the mixing with operators
containing quark fields. We construct the evolution equation for
corresponding correlator and because of its complexity we are forced
to find an effective approximation to it. By making use of the conformal
symmetry the problem is reduced finally to a quantum mechanical
problem for particles with a conformal invariant pair-wise interaction
and it turns out that the corresponding total Hamiltonian is a sum of an 
integrable piece equivalent to the Heisenberg spin chain of spin $s = 
- \frac{3}{2}$ and an addendum which breaks the integrability but still 
can be treated as a small perturbation for a large part of the spectrum. 
However, it plays its crucial r\^ole for the generation of a fine 
structure of lowest energy levels. Next, we address the non-singlet
quark-gluon-quark sector which in the limit of infinite number of
colours is shown to be identical to an inhomogeneous spin chain
with non-periodic boundary conditions. In both cases, gluon and
quark-gluon problems, we are entirely interested in the polynomial
solutions of the lattice models, which play an exceptional and
distinguished r\^ole for physics being implied by the OPE, since each
polynomial will correspond to a multiplicatively renormalizable
local operator. However, non-polynomial solutions are equally
interesting albeit their physical relevance within the present context
remains obscure presently. Finally, we summarize.

\section{Gluonic sector.}

In this section we address the question of the diagonalization of the
purely gluonic twist-three evolution equation for a correlation function 
which contributes to $g_2$. We will do it in parallel with an operator 
$\cT$ which possesses exactly solvable interaction and thus is very 
insightful for the realistic case. Thus, we introduce\footnote{The
$f_{abc}$-coupling in $\cT$-operator gives identical zero provided the
total derivatives are irrelevant like for the forward matrix elements
considered presently. Obviously, due to negative $C$-parity the 
$d_{abc}$-coupling does not contribute to the deep inelastic scattering 
cross section with an electromagnetic probe.}
\begin{equation}
\label{PositionSpace}
\left\{
\!\!
\begin{array}{c}
\cG_\alpha \\
\cT_{\alpha\beta\gamma}
\end{array}
\!\!
\right\}
(\kappa_1, \kappa_2, \kappa_3) = \g
\left\{
\!
\begin{array}{c}
f_{abc}\, g_{\alpha\nu}^\perp g_{\mu\rho}^\perp \\
d_{abc}\, \tau_{\alpha\beta\gamma;\mu\nu\rho}^\perp
\end{array}
\!
\right\}
G^{a \perp}_{+ \mu} ( \kappa_3 n )
G^{b \perp}_{+ \nu} ( \kappa_2 n )
G^{c \perp}_{+ \rho} ( \kappa_1 n ),
\end{equation}
where the totally symmetric, w.r.t.\ the independent permutation of
$\{ \alpha, \beta, \gamma\}$ and $\{ \mu, \nu, \rho\}$ indices, tensor 
is\footnote{Here the action of the operator of cyclic permutation
$\cP$ is defined as $\cP (1, 2, 3) f (1, 2, 3) \equiv f (1, 2, 3) +
f (2, 3, 1) + f (3, 1, 2)$.} $\tau_{\alpha\beta\gamma;\mu\nu\rho} \equiv
\cP (\alpha,\beta,\gamma) g_{\mu\alpha}^\perp
( g_{\nu\beta}^\perp g_{\rho\gamma}^\perp
+ g_{\nu\gamma}^\perp g_{\rho\beta}^\perp ) - \frac{1}{2}
\cP (\mu,\nu,\rho) g_{\mu\nu}^\perp ( g_{\alpha\beta}^\perp
g_{\rho\gamma}^\perp + g_{\alpha\gamma}^\perp g_{\rho\beta}^\perp +
g_{\beta\gamma}^\perp g_{\rho\alpha}^\perp )$.
The momentum fraction space functions are
\begin{equation}
\label{FractionSpace}
F (x_1, x_3) =
\int \frac{d \kappa_1}{2 \pi} \frac{d \kappa_3}{2 \pi}
e^{i x_1 \kappa_1 - i x_3 \kappa_3}
\langle h |
\cF (\kappa_1, 0, \kappa_3)
| h \rangle ,
\end{equation}
with $\cF = \cG, \cT$ and possess the symmetry properties $G (x_1, x_3)
= - G (- x_3, -x_1)$, $T (x_1, x_3) = T (- x_3, -x_1)$. They obey the
evolution equation
\begin{eqnarray}
\label{EvolutionEquation}
\frac{d}{d \ln Q^2} F (x_1, x_3)
= - \frac{\alpha_s}{2 \pi}
\int \prod_{i = 1}^{3} d x'_i \,
\delta \left( x'_1 - x'_3 + x'_2 \right)
\mbox{\boldmath$K$}^F \left( \{ x_i \} | \{ x'_i \} \right)
F (x'_1, x'_3) ,
\end{eqnarray}
where $x_2 = x_3 - x_1$.

Since the existing studies \cite{BukKurLip84,NonLoc} did not provide
an insight into the structure of the corresponding kernel
$\mbox{\boldmath$K$}^F$ we reanalyze the issue anew showing a simple
structure of the result which allows for a reduction of the problem
to a lattice model characterized by a Hamiltonian which consists of
two parts: exactly integrable piece equivalent to the generalized
Heisenberg spin chain and a one which violates this property and is
responsible for the formation of a fine structure of low lying
levels.

\subsection{Three-gluon evolution equation.}

To start with, note first of all that only `good' transverse components
of the fields enter the operators (\ref{PositionSpace}) and, therefore,
the latter belong to a special class of the so-called quasi-partonic
operators \cite{BukFroLipKur85}. They are distinguished by the property
that the renormalization does not move them outside of the class and that
at leading order the total evolution kernel is a linear combination of the
twist-two non-forward kernels in subchannels.

The pair-wise gluon evolution kernel can be decomposed as follows
into Lorentz tensors
\begin{eqnarray}
\label{ggKernel}
&&\!\!\!\!\!\!\!\!\!\!\!\!\!\! {^{gg}K^{ab;\mu\nu}_{a'b';\mu'\nu'}}
\left( x_1, x_2 | x'_1, x'_2 \right) \nonumber\\
&&= - \frac{\alpha_s}{2\pi}
\left\{
\frac{1}{2} g^\perp_{\mu\nu} g^\perp_{\mu'\nu'}
{^{gg} K^V}
+
\frac{1}{2} \epsilon^\perp_{\mu\nu} \epsilon^\perp_{\mu'\nu'}
{^{gg} K^A}
+
\tau^\perp_{\mu\nu;\rho\sigma} \tau^\perp_{\mu'\nu';\rho\sigma}
{^{gg} K^T}
\right\}^{ab}_{a'b'}
\left( x_1, x_2 | x'_1, x'_2 \right) ,
\end{eqnarray}
with $g^\perp_{\mu\nu} = g_{\mu\nu} - n_\mu n^\star_\nu
- n^\star_\mu n_\nu$, $\epsilon^\perp_{\mu\nu}
= \epsilon_{\mu\nu\rho\sigma} n^\star_\rho n_\sigma$ and
$\tau^\perp_{\mu\nu;\rho\sigma} = \frac{1}{2} \left( g^\perp_{\mu\rho}
g^\perp_{\nu\sigma}\right. + g^\perp_{\mu\sigma} g^\perp_{\nu\rho}
- \left. g^\perp_{\mu\nu} g^\perp_{\rho\sigma} \right)$; and colour
structures
\begin{equation}
\left\{ {^{gg} K^i} \right\}^{ab}_{a'b'}
= \frac{1}{N_c^2 - 1} \delta_{ab} \delta_{a'b'}
{^{gg} K^i}_{(1)}
+ \frac{1}{N_c} f_{abc} f_{a'b'c}
{^{gg} K^i}_{(8_A)}
+ \frac{N_c}{N_c^2 - 4} d_{abc} d_{a'b'c}
{^{gg} K^i}_{(8_S)}
+ \cdots ,
\end{equation}
where the dots stand for the decuplet and $27$-plet contributions
which are irrelevant for our consequent discussion.

\begin{figure}[t]
\begin{center}
\vspace{-0.3cm}
\hspace{1cm}
\mbox{
\begin{picture}(0,220)(270,0)
\put(0,-30){\inclfig{18}{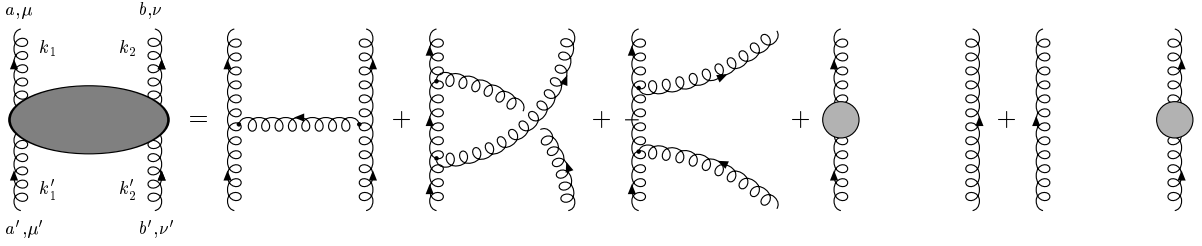}}
\end{picture}
}
\end{center}
\vspace{-5cm}
\caption{\label{gg-kernel} Diagrams (in the light-cone gauge) giving
rise to the gluon-gluon two-particle evolution kernel. The graph with
the crossed gluon line corresponds to the contact-type contribution
arising from the use of the equation of motion.}
\end{figure}

We can easily evaluate from diagrams in Fig.\ \ref{gg-kernel}
the kernels for the parity even and transversity channels
\begin{eqnarray}
f_{(c)}^{-1}\,
{^{gg}\! K^V_{(c)}} \left( x_1, x_2 | x'_1, x'_2 \right)
&=& k (x_1, x_2 | x'_1, x'_2)
+ 2 \frac{x_1 x_2 + x'_1 x'_2}{x'_1 x'_2}
\Theta^0_{111} (x_1, - x_2, x_1 - x'_1) \nonumber\\
&+& 2 \frac{x_1 x_2}{x'_1 x'_2}
\frac{x_1 x'_1 + x_2 x'_2}{(x_1 + x_2)^2}
\Theta^0_{11} (x_1, - x_2)
\pm ( x'_1 \leftrightarrow x'_2 ) , \\
f_{(c)}^{-1}\,
{^{gg}\! K^T_{(c)}} \left( x_1, x_2 | x'_1, x'_2 \right)
&=& k (x_1, x_2 | x'_1, x'_2)
\pm ( x'_1 \leftrightarrow x'_2 ) ,
\end{eqnarray}
with ``$+$'' sign for $f_{(8_S)} = f_{(1)}/2 = C_A/4$ and ``$-$'' sign
for $f_{(8_A)} = C_A/4$. For the parity odd sector we have
\begin{eqnarray}
f_{(c)}^{-1}
{^{gg}\! K^A_{(c)}} \left( x_1, x_2 | x'_1, x'_2 \right)
&=& k (x_1, x_2 | x'_1, x'_2)
+ 2 \frac{x_1 x'_2 + x'_1 x_2}{x'_1 x'_2}
\Theta^0_{111} (x_1, - x_2, x_1 - x'_1) \nonumber\\
&+& 2 \frac{x_1 x_2}{x'_1 x'_2}
\Theta^0_{11} (x_1, - x_2)
\mp ( x'_1 \leftrightarrow x'_2 ) , \nonumber
\end{eqnarray}
with ``$\pm$'' signs assigned vice versa to the previous case. The
generalized step functions look as follows
\begin{eqnarray*}
\Theta^0_{11} (x_1, x_2)
&=& \frac{\theta (x_1)}{x_1 - x_2} + \frac{\theta (x_2)}{x_2 - x_1},\\
\Theta^0_{111} (x_1, x_2, x_3)
&=& \frac{x_1 \theta (x_1)}{(x_2 - x_1)(x_1 - x_3)}
+ \frac{x_2 \theta (x_2)}{(x_1 - x_2)(x_2 - x_3)}
+ \frac{x_3 \theta (x_3)}{(x_1 - x_3)(x_3 - x_2)} .
\end{eqnarray*}
We have introduced the notation
\begin{eqnarray}
\label{IntegKer}
&&\!\!\!\!\!\!\!\! k (x_1, x_2 | x'_1, x'_2) \\
&&=
\frac{x_1}{x'_1}
\left[
\frac{x_1}{x_1 - x'_1} \Theta^0_{11} (x_1, x_1 - x'_1)
\right]_+
+
\frac{x_2}{x'_2}
\left[
\frac{x_2}{x_2 - x'_2} \Theta^0_{11} (x_2, x_2 - x'_2)
\right]_+
+ \left( \frac{1}{2} \frac{\beta_0}{C_A} + 2 \right)
\delta (x_1 - x'_1) , \nonumber
\end{eqnarray}
for the part which appears in all kernels.

\begin{figure}[t]
\begin{center}
\vspace{-0.3cm}
\hspace{4cm}
\mbox{
\begin{picture}(0,220)(270,0)
\put(0,-30){\inclfig{18}{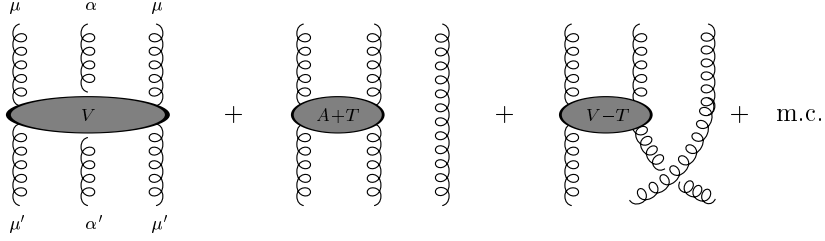}}
\end{picture}
}
\end{center}
\vspace{-5.0cm}
\caption{\label{Gtotal} Structure of the total evolution kernel for the
three-gluon correlation functions $G (x_1, x_3)$. The symbol m.c. stands
for the mirror symmetrical contributions of the last two diagrams.}
\end{figure}

Introducing the conventions
\begin{equation}
\widehat{K} ( x_i, x_j | x'_i, x'_j )
=
K ( x_i, x_j | x'_i, x'_j )
\delta ( x_i + x_j - x'_i - x'_j ),
\end{equation}
and $K^{i \pm j} \equiv \frac{1}{2} \left( K^i \pm K^j \right)$, we can
contract the generalized kernels (\ref{ggKernel}) with the tensor
structure of the operators in question (\ref{PositionSpace}) and
construct easily the total evolution equation for the three-gluon
quasi-partonic correlator $G$ (see Fig.\ \ref{Gtotal} for its
pictorial representation)
\begin{eqnarray}
\label{totGkernel}
{\mbox{\boldmath$K$}^G}
\!\!\!&=&\!\!\!
2\, {^{gg}\widehat{K}^{A + T}_{(8_A)}} (x_1, x_2 | x'_1, x'_2)
+ 2\, {^{gg}\widehat{K}^{A + T}_{(8_A)}} (x_2, - x_3 | x'_2, - x'_3)
\nonumber\\
&+&\!\!\! {^{gg}\widehat{K}^V_{(8_A)}} (x_1, - x_3 | x'_1, - x'_3)
- {^{gg}\widehat{K}^{V - T}_{(8_A)}} (x_1, x_2 | x'_1, - x'_3)
- {^{gg}\widehat{K}^{V - T}_{(8_A)}} (x_2, - x_3 | x'_1, - x'_3)
\nonumber\\
&-&\!\!\! {\scriptstyle\frac{1}{4}}\beta_0
\delta (x_1 - x'_1) \delta (x_3 - x'_3) ,
\end{eqnarray}
and for the $T$
\begin{eqnarray}
\label{totTkernel}
{\mbox{\boldmath$K$}^T}
\!\!\!&=&\!\!\!
{^{gg}\widehat{K}^T_{(8_S)}} (x_1, x_2 | x'_1, x'_2)
+ {^{gg}\widehat{K}^T_{(8_S)}} (x_2, - x_3 | x'_2, - x'_3)
+ {^{gg}\widehat{K}^T_{(8_S)}} (x_1, - x_3 | x'_1, - x'_3)
\nonumber\\
&-&\!\!\! {\scriptstyle\frac{1}{4}}\beta_0
\delta (x_1 - x'_1) \delta (x_3 - x'_3) .
\end{eqnarray}

Defining the moments as
\begin{equation}
\label{Moments}
F_j^J = \int d x_1 d x_3 x_1^j x_3^{J - j} F (x_1, x_3),
\end{equation}
with the properties $G_j^J = - ( - 1)^J G_{J - j}^J$, $T_j^J = ( - 1)^J
T_{J - j}^J$ we can write, given the results of Eqs.\ (\ref{totGkernel})
and (\ref{totTkernel}), the evolution equation for the moments, i.e.
for the local operators $F_j^J \propto \g G_{+ \perp} (i \partial_+)^j
G_{+ \perp} (i \partial_+)^{J - j} G_{+ \perp}$, in the form:
\begin{equation}
\frac{d}{d \ln Q^2} F_j^J = - \frac{\alpha_s}{4 \pi}
\sum_{k = 0}^{J}
{\mbox{\boldmath${\mit\Gamma}$}_{jk}^F} (J)
F_k^J ,
\qquad
{\mbox{\boldmath${\mit\Gamma}$}_{jk}^F} (J)
= C_A {\mbox{\boldmath${\gamma}$}_{jk}^F} (J)
+ \beta_0 \delta_{jk} ,
\end{equation}
with the anomalous dimension matrices
\begin{eqnarray}
\label{ADmatrixG}
{\mbox{\boldmath${\gamma}$}_{jk}^G} (J)
\!\!\!&=&\!\!\!
\delta_{jk} \Bigg\{
3 \psi (j + 3) - 3 \psi (1)
+ 2 \frac{(- 1)^{J - j}}{(j + 1)_4}
\left( (j + 1) C_{J + 2}^j + 3 (j + 2) C_{J + 1}^j \right)
- \frac{1}{J + 2}
\nonumber\\
&&\quad
- 2 \frac{(j + 1)(J - j + 1)}{(J + 2)_3}
- 2 \frac{1 + (- 1)^j }{(j + 2)_2}
\Bigg\} \nonumber\\
&+&\!\!\! \theta_{k, j + 1} \Bigg\{
2 (-1)^{k - j}
\frac{C_{J - j + 1}^{J - k + 1}}{C_{k + 1}^{j + 1}}
\left(
\frac{k - j}{(J + 4)(k + 2)} \right.
+
\frac{k - j}{(J + 2)(J - j + 1)}
- \frac{(j + 1)(J - k + 1)}{(J + 2)_3}
\nonumber\\
&&\quad
- \left. \frac{(k + 1)(J - j + 1)}{(J + 2)_3}
- \frac{1}{J + 2}
\right)
- 2 \frac{J - k + 1}{(J - j + 1)_3}
\left( 1 + (- 1)^{J - k} C_{J - j + 2}^{J - k + 2} \right) \nonumber\\
&&\quad
+ 2 \frac{(- 1)^{J - k}}{(j + 1)_4}
\left(
(j + 1) \left( C_{J + 2}^{k} - C_{J - j}^{J - k + 2} \right)
+ (j + 2) \left( 3 C_{J + 1}^{k} - (j + 4) C_{J - j}^{J - k + 1} \right)
\right) \nonumber\\
&&\quad
+ 2 \frac{(- 1)^k}{(J - j + 1)_4}
\left(
(J - j + 1) C_{J + 2}^{k + 2} + 3 (J - j + 2) C_{J + 1}^{k + 1}
\right)
\nonumber\\
&&\quad
- \frac{1}{k - j}
\left( (-1)^{k - j}
\frac{C_{J - j}^{J - k}}{C_{k + 2}^{j + 2}}
+ \frac{(J - k + 1)_2}{(J - j + 1)_2}
\right)
\Bigg\} + {j \rightarrow J - j \choose k \rightarrow J - k} ,
\end{eqnarray}
and\footnote{To get these anomalous dimensions we assumed the gluons
being different so that there is no symmetrization in the pair-wise
kernels. Identity of gluons will just pick up eigenstates with a
definite symmetry. With the matrix (\ref{ADmatrixT}) at hand we have 
the states with all possible symmetries.}
\begin{eqnarray}
\label{ADmatrixT}
{\mbox{\boldmath${\gamma}$}_{jk}^T} (J)
\!\!\!&=&\!\!\!
\delta_{jk} \left\{
3 \psi (j + 3) - 3 \psi (1)
\right\} \nonumber\\
&-&\!\!\! \theta_{k, j + 1} \frac{1}{k - j}
\left\{
(- 1)^{k - j}
\frac{C_{J - j}^{J - k}}{C_{k + 2}^{j + 2}}
+ \frac{(J - k + 1)_2}{(J - j + 1)_2}
\right\} + {j \rightarrow J - j \choose k \rightarrow J - k} ,
\end{eqnarray}
where we introduced the Pochhammer symbol $( j )_\ell = \frac{\Gamma
(j + \ell)}{\Gamma (j)}$, the binomial coefficients $C_J^j = \frac{
\Gamma (J + 1)}{\Gamma (j + 1) \Gamma (J - j + 1)}$ and the step
functions $\theta_{j,k} = \{1,\ \mbox{if}\ j \geq k;\ 0,\ \mbox{if}\
j < k \}$. Our mixing matrix (\ref{ADmatrixG}) differs from the one
obtained in Ref.\ \cite{BukKurLip84} since the evolution equation was
written there for a different quantity, i.e.\ a linear combination of
$G$-functions considered here (\ref{PositionSpace},\ref{FractionSpace}).
But both matrices have indeed identical eigenvalues with different
eigenvectors.

\subsection{Reduction to 1D lattice model.}

\begin{figure}[t]
\unitlength1mm
\begin{center}
\vspace{-1cm}
\hspace{0cm}
\begin{picture}(100,155)(0,0)
\put(0,60){\insertfig{9}{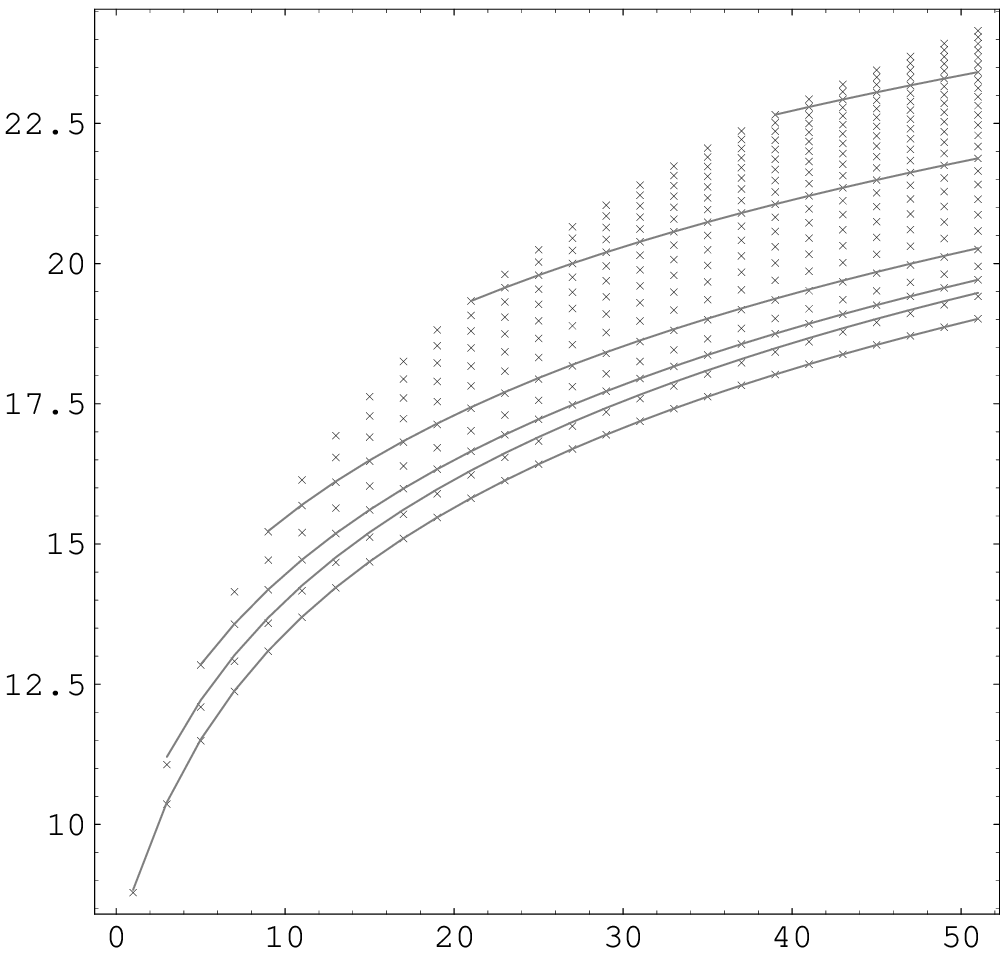}}
\put(46,55){$J$}
\put(-10,100){\rotate{$\cE (J)$}}
\end{picture}
\end{center}
\vspace{-6cm}
\caption{\label{ExactVsAppr} The exact energy spectrum (\ref{ADmatrixG})
versus the selected trajectories generated by the approximate formula
(\ref{SimpleADmatrix}).}
\end{figure}

The key observation is that not all of the parts of the anomalous
dimension matrix (\ref{ADmatrixG}) play an equally important r\^ole
in the generation of the energy spectrum. Note first that due to the
antisymmetry of the three-particle correlator $G (x_1, x_3)$, w.r.t.\ 
the interchange of the momentum fractions of the first and last gluon, 
the leading rightmost $\frac{1}{j}$-singularity in the $j$-plane of 
the vector kernel disappears. For the formation of the upper part of 
the spectrum $j,k \sim J$ only $[1/(x_i - x'_i)]_+$-distributions are 
relevant while the fine structure of low levels is generated by the 
$K^A + K^T$-part of the total kernel. Making use of these observations 
we can deduce a simplified evolution kernel which reproduces with a 
good accuracy the exact one. Namely, the simplified anomalous dimension 
matrix reads
\begin{eqnarray}
\label{SimpleADmatrix}
{\mbox{\boldmath${\gamma}$}_{jk}} (J)
\!\!\!&=&\!\!\!
\delta_{jk} \left\{
3 \psi (j + 3) - 3 \psi (1)
- \frac{4}{(j + 2)_2}
\right\} \\
&-&\!\!\! \theta_{j - 1, k}
\left\{
\frac{1}{k - j}
\left(
(- 1)^{k - j}
\frac{C_{J - j}^{J - k}}{C_{k + 2}^{j + 2}}
+ \frac{(J - k + 1)_2}{(J - j + 1)_2}
\right)
+ 4\frac{J - k + 1}{(J - j + 1)_3}
\right\}
+ {j \rightarrow J - j \choose k \rightarrow J - k} , \nonumber
\end{eqnarray}
and its eigenvalues coincide with the ones of Eq.\ (\ref{ADmatrixG}) with 
a high precision, see Fig.\ \ref{ExactVsAppr}. This anomalous dimension 
matrix corresponds to the kernel
\begin{eqnarray}
\label{SimplifiedKernel}
{\mbox{\boldmath$K$}} \!\!\!&=&\!\!\!
\frac{C_A}{2}
\Big\{
\widehat{k} (x_1, x_2 | x'_1, x'_2)
+ \widehat{k} (x_2, - x_3 | x'_2, - x'_3)
+ \widehat{k} (x_1, - x_3 | x'_1, - x'_3) \nonumber\\
&+& \widehat{\Delta} (x_1, x_2 | x'_1, x'_2)
+ \widehat{\Delta} (x_2, - x_3 | x'_2, - x'_3)
\Big\} - {\scriptstyle\frac{1}{4}}\beta_0
\delta (x_1 - x'_1) \delta (x_3 - x'_3) ,
\end{eqnarray}
with $k$ defined in Eq.\ (\ref{IntegKer}) and
\begin{equation}
\Delta (x_1, x_2 | x'_1, x'_2)
=
2 \frac{x_1 x'_2 + x'_1 x_2}{x'_1 x'_2}
\Theta^0_{111} (x_1, - x_2, x_1 - x'_1)
+ 2 \frac{x_1 x_2}{x'_1 x'_2}
\Theta^0_{11} (x_1, - x_2) .
\end{equation}

As is well known the tree level conformal invariance of QCD
\cite{MacSal69} allows for the diagonalization of the pair-wise
evolution kernels in subchannels by means of the Jacobi polynomials
$P^{(\nu_b, \nu_a)}_j$ \cite{EfrRad79}-\cite{BelMul99a} and
express the two-particle Hamiltonians as a function of corresponding
Casimir operators. Hence, we find
\begin{equation}
\int dx_1 dx_2 P^{\left( 2, 2 \right)}_j
\left( \frac{x_1 - x_2}{x_1 + x_2} \right) \,
\left\{ {\widehat{k} \atop \widehat{\Delta}} \right\}
\left( x_1, x_2 | x'_1, x'_2 \right)
= \left\{
{ 2 \psi (j + 3) - 2 \psi (1) + \frac{1}{2}\frac{\beta_0}{C_A}
\atop
- \frac{4}{(j + 2)(j + 3)} }
\right\}
P^{\left( 2, 2 \right)}_j
\left( \frac{x'_1 - x'_2}{x'_1 + x'_2} \right),
\end{equation}
where $j$ is a conserved quantum number related to the conformal spin
of the composite two-particle conformal operator $\cO_{jk} = \phi_b
(i \partial_a + i \partial_b)^k P^{(\nu_b, \nu_a)}_j \left( (\partial_a
- \partial_b)/(\partial_a + \partial_b) \right) \phi_a$, composed from
elementary fields of conformal weight $\nu = d + s - 1$ ($= 2$ for gluons).
Namely, the eigenvalues of the pair-wise Casimir operator of the collinear
conformal algebra $su(1,1)$, $\mbox{\boldmath$\hat J$}^2_{ab}$, on these
states are $J_{ab}(J_{ab} - 1)$ where $J_{ab} = j + \frac{1}{2}(\nu_a
+ \nu_b + 2)$. Therefore, we can substitute the original problem of the
diagonalization of the kernel (\ref{SimplifiedKernel})
\begin{equation}
{\mbox{\boldmath$K$}} \equiv
\frac{C_A}{2} \, {\mbox{\boldmath$H$}}
\left( \{ x_i \} | \{ x'_i \} \right)
+ \frac{\beta_0}{2}
\delta(x_1 - x'_1) \delta (x_3 - x'_3) ,
\end{equation}
by a quantum mechanical problem of the diagonalization of the
Hamiltonian, ${\mbox{\boldmath$H$}} \to \cH$,
\begin{equation}
\label{PertHamilton}
{\cH} = {\cH}_0 + {\cV},
\end{equation}
where
\begin{equation}
{\cH}_0 = h_{12} + h_{23} + h_{31}, \qquad
{\cV} = v_{12} + v_{23},
\end{equation}
with the pair-wise interaction
\begin{equation}
\label{QCDlocalHam}
h_{ab} = 2 \psi \left( \hat J_{ab} \right) - 2 \psi (1) , \qquad
v_{ab} = - \frac{4}{\mbox{\boldmath$\hat J$}^2_{ab}} .
\end{equation}
Here the operator $\hat J_{ab}$ is defined as a formal solution of the
equation $\mbox{\boldmath$\hat J$}^2_{ab} = \hat J_{ab} (\hat J_{ab}
- 1)$ and possesses the eigenvalues $J_{ab}$. Obviously, $[\cH,
\mbox{\boldmath$\hat J$}^2]_- = 0$ where $\mbox{\boldmath$\hat J$}^2
= \mbox{\boldmath$\hat J$}^2_{12} + \mbox{\boldmath$\hat J$}^2_{23}
+ \mbox{\boldmath$\hat J$}^2_{31} - 9/4$ is the total Casimir
operator of the gluonic system and, therefore, it is the integral of
motion. The spectrum of $\cH_0$ part corresponds to the anomalous
dimensions of the operator $\cT$.

The explicit form of the operators $\mbox{\boldmath$\hat J$}^2_{ab}$
depends on a basis where their action is defined. For our purposes
we find it convenient to use the space spanned by the elements
$\theta^k \equiv \frac{\partial^k_+ \phi}{\Gamma (k + \nu + 1)}$
\cite{Bel99a,Bel99b}. Then, it is easy to read off their form from the
action of the generators of the collinear conformal group on the
elementary field operators. Finally, we have for the step-up, the
step-down and the grade operators the representation
\begin{equation}
\hat J^+ = (\nu + 1) \theta + \theta^2 \frac{\partial}{\partial\theta},
\quad
\hat J^- = \frac{\partial}{\partial\theta},
\quad
\hat J^3 = \frac{1}{2}(\nu + 1) + \theta \frac{\partial}{\partial\theta} .
\end{equation}
The quadratic Casimir operator is given by ${\mbox{\boldmath$\hat J$}}^2
= \hat J^3 ( \hat J^3 - 1 ) - \hat J^+ \hat J^-$ and equals
${\mbox{\boldmath$\hat \jmath$}}^2 \equiv \frac{1}{4} ( \nu^2 - 1 )$
for a one-particle state. The two-particle Casimir which enters Eq.\
(\ref{QCDlocalHam}) is
\begin{equation}
\label{TwoPartCasimir}
{\mbox{\boldmath$\hat J$}}^2_{ab}
= - \theta_{ab}^{1 - ( \nu_a + \nu_b )/2}
\partial_a \partial_b
\theta_{ab}^{1 + ( \nu_a + \nu_b )/2}
+ \frac{1}{2} ( \nu_b - \nu_a ) \theta_{ab}
\left( \partial_a + \partial_b \right) .
\end{equation}

\subsection{Isotropic Heisenberg magnet.}

In this section we show that the Hamiltonian $\cH_0$ possesses, apart
from $\mbox{\boldmath$\hat J$}^2$, an extra integral of motion which
makes the former exactly integrable. We will find that it corresponds to
the generalized homogeneous XXX chain \cite{IzeKor81} with sites of equal
spin $s = - \frac{\nu + 1}{2} = - \frac{3}{2}$. To demonstrate this we
use the formalism of the Quantum Inverse Scattering Method
\cite{SklTakFad79,Fad80,FadTak79,KorBogIze93} and consider the
lattice with equal conformal weights on each site $\nu_\ell = \nu$.
To prove the integrability of the Schr\"odinger equation
\begin{equation}
\cH_0 {\mit\Psi}_0 = \cE_0 {\mit\Psi}_0,
\end{equation}
let us define a matrix $R_{a,b}$ acting on the product of the vector
space of the site $V_a$ and an auxiliary space $V_b$. It satisfies
the Yang-Baxter equation \cite{Fad80,FadTak79,KorBogIze93,KulSkl82}
\begin{equation}
\label{YangBaxter}
R_{a,b} ( \lambda - \mu ) R_{c,a} ( \lambda ) R_{c,b} ( \mu )
=
R_{b,c} ( \mu ) R_{a,c} ( \lambda ) R_{a,b} ( \lambda - \mu ).
\end{equation}
The solution to this equation for the case when the dimensions of the
quantum and auxiliary spaces coincide and which is of the most interest
for us is given by \cite{KulResSkl81,TarTakFad83,Fad95}
\begin{equation}
\label{YBbundle}
R_{a,b} (\lambda) = f(\nu, \lambda) P_{ab}
\frac{\Gamma (\hat J_{ab} + \lambda) \Gamma (\nu + 1 - \lambda)}{
\Gamma (\hat J_{ab} - \lambda) \Gamma (\nu + 1 + \lambda)} .
\end{equation}
It is defined up to an arbitrary c-number function $f (\nu, \lambda)$.
In the case when the auxiliary space is two-dimensional the $R$-matrix
gives the Lax operator
\begin{equation}
L_a (\lambda) \equiv
R_{a,\frac{1}{2}} \left( \lambda - \frac{1}{2} \right)
= \lambda \1 + \sigma^i \hat J^i_a ,
\end{equation}
where $\mbox{\boldmath$\hat J$} = (\hat J^1, \hat J^2, \hat J^3)$ with
$\hat J^1 = \frac{1}{2} \left( \hat J^- - \hat J^+ \right)$ and
$\hat J^2 = \frac{i}{2} \left( \hat J^- + \hat J^+ \right)$.

Following a standard procedure we define the auxiliary and the fundamental
monodromy matrices
\begin{equation}
T_{\frac{1}{2}} (\lambda)
= L_{a_1} (\lambda) L_{a_2} (\lambda) L_{a_3} (\lambda),
\quad
T_b (\lambda)
= R_{a_1,b} (\lambda) R_{a_2,b} (\lambda) R_{a_3,b} (\lambda) ,
\end{equation}
satisfying the $RTT$-equation which is Eq.\ (\ref{YangBaxter})
but with two $R$-matrices replaced by the monodromies. It means that
$R (\lambda - \mu)$ intertwines the two possible co-multiplications of
the $T$-matrices. This immediately leads to the commutation relations
\begin{equation}
\label{Commutativity}
[t_{b_1} (\lambda_1), t_{b_2} (\lambda_2)]_- = 0
\end{equation}
between the transfer matrices,
\begin{equation}
t_{b} (\lambda) = {\rm tr}_b T_{b} (\lambda),
\end{equation}
which act on the total space of the spin chain
$\OO_{\ell = 1}^3 V_{a_\ell}$.

The fundamental transfer matrix gives us the total Hamiltonian of the
chain \cite{KulSkl82,KulResSkl81,TarTakFad83}
\begin{equation}
\cH_0 = \left.\frac{d}{d \lambda}\right|_{\lambda = 0}
\ln t_b (\lambda)
= h_{a_1, a_2} + h_{a_2, a_3} + h_{a_3, a_1},
\end{equation}
with the two-site Hamiltonians
\begin{equation}
\label{TwoSiteHamilton}
h_{a, b} = R_{a, b} (0) R'_{a, b} (0) =
2 \psi \left( \hat J_{ab} \right) - 2 \psi (1) ,
\end{equation}
coinciding with Eq.\ (\ref{QCDlocalHam}) for the one-particle conformal
weight $\nu = 2$ and provided $f (\nu, \lambda) = \frac{\Gamma (1 -
\lambda)}{\Gamma (1 + \lambda)} \frac{\Gamma (\nu + 1 + \lambda)}{
\Gamma (\nu + 1 - \lambda)}$. The expansion of the auxiliary transfer
matrix in the rapidity $\lambda$ gives
\begin{equation}
t_\frac{1}{2} (\lambda)
= 2 \lambda^3
+ \lambda \left(
{\mbox{\boldmath$\hat J$}}^2
- 3 {\mbox{\boldmath$\hat \jmath$}}^2
\right)
+ i \cQ,
\end{equation}
where the total conformal spin and the `hidden' charge are
\begin{equation}
{\mbox{\boldmath$\hat J$}}^2
= \left( \mbox{\boldmath$\hat J$}_1 + \mbox{\boldmath$\hat J$}_2
+ \mbox{\boldmath$\hat J$}_3 \right)^2,
\qquad
\cQ = 2 \epsilon_{ijk} \hat J^i_1 \hat J^j_2 \hat J^k_3 ,
\end{equation}
respectively. Note that $\cQ$ can be represented as a commutator of the
two-particle Casimir operators (\ref{TwoPartCasimir})
${\mbox{\boldmath$\hat J$}}^2_{ab} = \left( \mbox{\boldmath$\hat J$}_a
+ \mbox{\boldmath$\hat J$}_b \right)^2$ making use of the relation
$\epsilon_{ijk} \hat J^i_1 \hat J^j_2 \hat J^k_3 = \frac{i}{4}
[ {\mbox{\boldmath$\hat J$}}^2_{12} ,
{\mbox{\boldmath$\hat J$}}^2_{23} ]_-$ which allows to rewrite it
equivalently as $[ {\mbox{\boldmath$\hat J$}}^2_{12} ,
{\mbox{\boldmath$\hat J$}}^2_{23} ]_-
= [ {\mbox{\boldmath$\hat J$}}^2_{13} ,
{\mbox{\boldmath$\hat J$}}^2_{12} ]_-
= [ {\mbox{\boldmath$\hat J$}}^2_{23} ,
{\mbox{\boldmath$\hat J$}}^2_{13} ]_-$.

For the later convenience we introduce a convention for the eigenvalues
of the quadratic Casimir operator of the chain, namely,
\begin{equation}
\eta^2 \equiv \left( J + \frac{3}{2} (\nu + 1) \right)
\left( J + \frac{3}{2} (\nu + 1) - 1 \right) .
\end{equation}
For gluons at large $J$ we have $\eta = J + 4 + \cO (J^{- 1})$.

\subsection{Conformal basis and recursion relation.}
\label{RecursionRelationWKB}

The eigenfunction of the three-particle system can be found making
use of the existence of the additional non-trivial integral of motion
\begin{equation}
\label{QchargeEq}
\cQ {\mit\Psi}_0 = q {\mit\Psi}_0,
\end{equation}
where $q$ are the eigenvalues of $\cQ$.

The operator product expansion suggests that we must be primarily
interested in polynomial solutions of the above equation. We can
make further profit of conformal invariance of the system by looking
the solution to this equation in the form of expansion w.r.t.\ a
three-particle basis:
\begin{equation}
\label{EigenFunExpansion}
{\mit\Psi}_0 = \sum_{j = 0}^{J} {\mit\Psi}_j
\cP_{J;j} (\theta_1, \theta_3 | \theta_2) ,
\end{equation}
where the basis vectors
\begin{equation}
\label{BasisGGG}
\cP_{J;j} (\theta_1, \theta_3 | \theta_2)
= \varrho_j^{- 1}
\frac{(2 j + 5)}{(j + 3)_{j + 3}}
\frac{\Gamma (J + j + 6)}{\Gamma^{1/2} (2 J + 8)}
\theta_{13}^J \, \theta^{j - J}
{_2F_1} \left( \left.
{ j - J , j + 3 \atop 2j + 6 }
\right| \theta \right) ,
\quad\mbox{with}\quad
\theta \equiv \frac{\theta_{13}}{\theta_{23}} ,
\end{equation}
diagonalize simultaneously \cite{BraDerKorMan99,Bel99a,Bel99b}
the total ${\mbox{\boldmath$\hat J$}}^2$
and ${\mbox{\boldmath$\hat J$}}^2_{13}$ Casimir operators
\begin{eqnarray}
\label{DiagTot}
{\mbox{\boldmath$\hat J$}}^2
\cP_{J; j} ( \theta_1, \theta_3 | \theta_2 )
\!\!\!&=&\!\!\!
\left( J + \frac{9}{2} \right)\left( J + \frac{7}{2} \right)
\cP_{J; j} ( \theta_1, \theta_3 | \theta_2 ) , \\
\label{Diag12}
{\mbox{\boldmath$\hat J$}}^2_{13}
\cP_{J; j} ( \theta_1, \theta_3 | \theta_2 )
\!\!\!&=&\!\!\!
\left( j + 2 \right) \left( j + 3 \right)
\cP_{J; j} ( \theta_1, \theta_3 | \theta_2 ) ,
\end{eqnarray}
and are normalized to unity
$\langle {\cal P}_{J';j'} ( \theta_1, \theta_2 | \theta_3 )|
{\cal P}_{J;j} ( \theta_1, \theta_2 | \theta_3 ) \rangle =
\delta_{J'J} \delta_{j'j}$ w.r.t. $SU(1,1)$ invariant scalar product
\begin{equation}
\label{ScalarProduct}
\langle \chi_2 ( \theta_1, \theta_2, \theta_3 ) |
\chi_1 ( \theta_1, \theta_2, \theta_3 ) \rangle
= \int\limits_{ |\theta_\ell| \leq 1 }
\prod_{\ell = 1}^{3}
\frac{d \theta_\ell d \bar\theta_\ell}{2 \pi i}
( 1 - \theta_\ell \bar\theta_\ell )^{\nu_\ell - 1}
\
\chi_2 ( \bar\theta_1, \bar\theta_2, \bar\theta_3 )
\chi_1 ( \theta_1, \theta_2, \theta_3 ) .
\end{equation}

The Hamiltonian $\cH_0$ of the system exhibits cyclic permutation
symmetry of the sites of the chain. Therefore, its eigenfunctions
${\mit\Psi}_0$ could change at most their phase under this transformation
\begin{equation}
P {\mit\Psi}_0 (\theta_1, \theta_3 | \theta_2)
\equiv
{\mit\Psi}_0 (\theta_2, \theta_1 | \theta_3)
= e^{i \varphi}
{\mit\Psi}_0 (\theta_1, \theta_3 | \theta_2) .
\end{equation}
Then
\begin{equation}
\theta
\stackrel{P}{\to}
1 - \frac{1}{\theta}
\stackrel{P}{\to}
\frac{1}{1 - \theta} ,
\end{equation}
for the ``anharmonic" ratio of the conformal basis. Since the triple
action of $P$ leads to the system at its initial state, the eigenvalues
of the operator are cubic roots of unity
\begin{equation}
P^3 = 1, \qquad \varphi = 0 ,\ \frac{2}{3} \pi ,\ \frac{4}{3} \pi .
\end{equation}
Since this phase appears as a result of the cyclic shift of the chain
sites it has an obvious interpretation of the momentum of the lattice.
Obviously, it is the function of the conserved charges and is expressed
through the equality
\begin{equation}
\varphi = \arg
\frac{\sum_{j = 0}^{J} (- i)^j (2 j + 5) {\mit\Upsilon}_j}{
\sum_{j = 0}^{J} i^j (2 j + 5) {\mit\Upsilon}_j} ,
\end{equation}
with ${\mit\Upsilon}_j$ related to the expansion coefficients
${\mit\Psi}_j$ via Eq.\ (\ref{Relation}). It follows from this
definition that $\varphi (- q) = - \varphi (q)$.

The cyclic permutation transforms different three-particle bases
and since each of them corresponds just to a different quantum
mechanical addition of three spins they are related by the conventional
$6j$-symbols as
\begin{equation}
\label{RacahRelation}
\cP_{J;j} (\theta_2, \theta_1 | \theta_3)
= \sum_{k = 0}^{J} W_{jk} (J)
\cP_{J;j} (\theta_1, \theta_3 | \theta_2) .
\end{equation}
The Racah coefficients fulfill the following properties:
$\sum_{\ell = 0}^{J} W_{j \ell} W_{k \ell} = \delta_{jk}$ which is a
consequence of the completeness condition $\sum_{j = 0}^{J} | \cP_{J;j}
\rangle \langle \cP_{J;j} | = 1$; $\sum_{\ell, m = 0}^{J} W_{j \ell}
W_{\ell m} W_{m k} = \delta_{jk}$ coming from $P^3 = 1$; and
$W_{jk} = (- 1)^{j + k} W_{kj}$ reflecting the relation
$P^2 = P_{13} P P_{13}$.

Introducing new conventions
\begin{equation}
\label{Relation}
{\mit\Psi}_j = (- i)^j \varrho_j {\mit\Upsilon}_j ,
\end{equation}
with
\begin{equation}
\varrho_j \equiv
\left[
\frac{(j + 1)_4 (J + j + 6)_2 (J - j + 1)_2}{2j + 5}
\right]^{- 1/2} ,
\end{equation}
we can deduce from Eq.\ (\ref{QchargeEq}) the recursion relation
for the expansion coefficients ${\mit\Upsilon}_j$
\begin{equation}
\label{RecRel}
q {\mit\Upsilon}_j
= a_j {\mit\Upsilon}_{j + 1} + b_j {\mit\Upsilon}_{j - 1},
\end{equation}
where
\begin{equation}
a_j = \frac{(j + 1)(j + 3)}{2 (2j + 5)} (J + j + 6) (J - j + 2),
\qquad
b_j = \frac{(j + 2)(j + 4)}{2 (2j + 5)} (J + j + 7) (J - j + 1) ,
\end{equation}
and satisfying the boundary conditions
\begin{equation}
{\mit\Upsilon}_{-1} = {\mit\Upsilon}_{J + 1} = 0 .
\end{equation}
Obviously, the coefficients ${\mit\Upsilon}_j$ are related for
positive and negative values of the eigenvalues of integral of
motion due to $P_{13} \cQ P_{13}^{- 1} = - \cQ$ as
\begin{equation}
\label{SymmUpsilon}
{\mit\Upsilon}_j (- q) = (- 1)^j {\mit\Upsilon}_j (q) .
\end{equation}
And since $P_{13} \cP_{J;j} (\theta_1, \theta_3 | \theta_2) = (- 1)^j
\cP_{J;j} (\theta_1, \theta_3 | \theta_2)$ we conclude from here
that $P_{13} {\mit\Psi}_0 (q) = {\mit\Psi}_0 (- q)$.

\subsection{Conserved charge.}

The exact analytical solution based on the Eq.\ (\ref{RecRel}) is
difficult. We, therefore, perform our analysis for large conformal
spin $J$.

There exists, however, the exceptional point $q = 0$ when the exact
solution can be easily found since Eq.\ (\ref{RecRel}) is reduced to
a two-term relation. Due to Eq.\ (\ref{SymmUpsilon}) we have
${\mit\Upsilon}_{2k + 1} (0) \equiv 0$ with $k = 0, 1, \dots, J/2$
and from the condition of fulfilling the boundary condition
${\mit\Upsilon}_{J + 1} = 0$ we conclude that this solution exists only
for even $J$. For even $j \equiv 2 k$ we get
\begin{equation}
\label{EigenfunQzero}
{\mit\Upsilon}_{2k} (0)
= (- 1)^k {\mit\Upsilon}_0 (0)
\prod_{\ell = 1}^{k} \frac{b_{2 \ell - 1}}{a_{2 \ell - 1}}
= \frac{(- 1)^k}{3 \cdot 4^k}
\frac{(2k + 1)!! (2k + 3)!!}{k! (k + 1)!}
\frac{\left( \frac{J + 2}{2} - k \right)_k
\left( \frac{J + 8}{2} \right)_k}{
\left( \frac{J + 3}{2} - k \right)_k
\left( \frac{J + 7}{2} \right)_k} \,
{\mit\Upsilon}_0 (0) .
\end{equation}

\begin{figure}[t]
\unitlength1mm
\begin{center}
\vspace{-1cm}
\hspace{0cm}
\begin{picture}(100,155)(0,0)
\put(0,60){\insertfig{9}{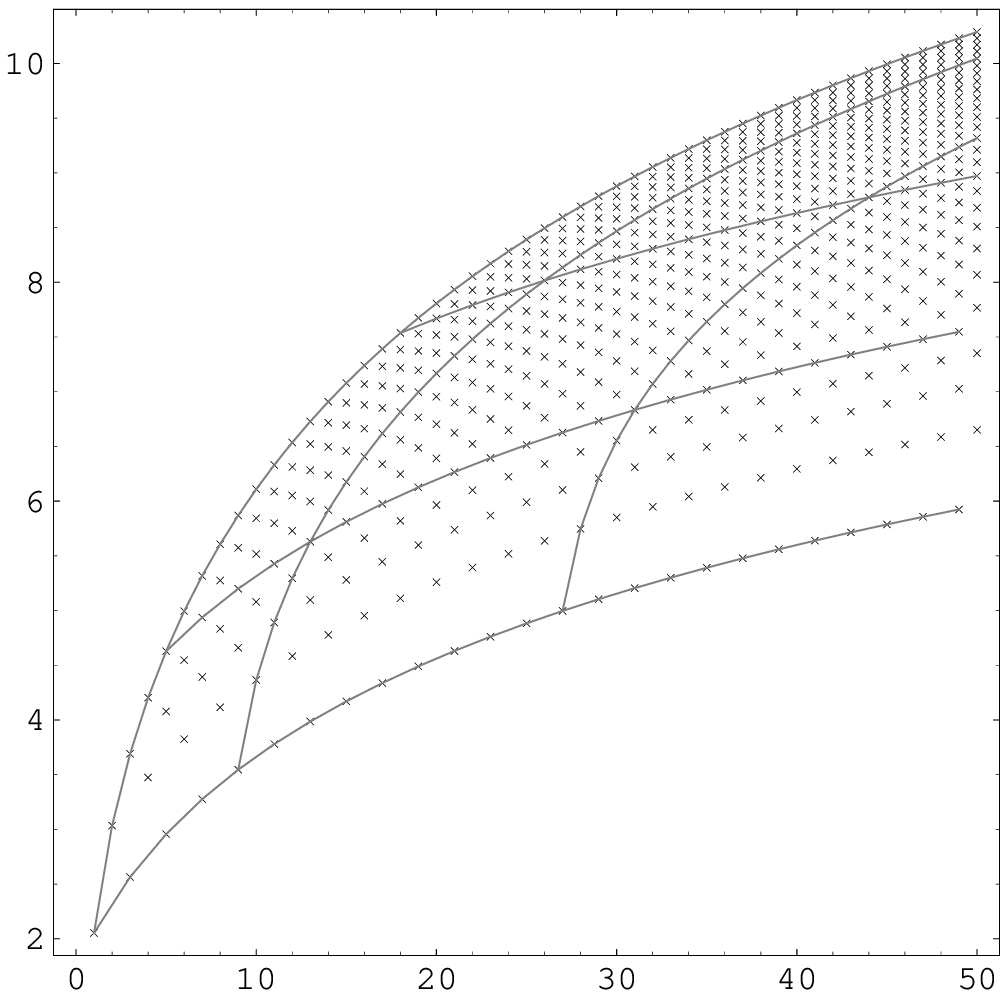}}
\put(46,55){$J$}
\put(-10,102){\rotate{$\ln |q (J)|$}}
\end{picture}
\end{center}
\vspace{-6cm}
\caption{\label{AbsQ} The log of the absolute value of the numerically
diagonalized conserved charge $q$ with $q = 0$ trajectory being
removed. The two possibilities of the description are clearly seen: from
the top by the trajectories which behave at large $J$ as $q \propto J^3$;
and from the bottom with the behaviour $q \propto J^2$.}
\end{figure}

Now we proceed with the rest of the spectrum. From the condition
$q^2 \leq 4 a_j b_j$ of the existence of solutions to the recursion
relation (\ref{RecRel}) we can easily deduce the critical values of the
conserved charge for asymptotical conformal spins $J \to \infty$ to be
\begin{equation}
- \frac{J^3}{3 \sqrt{3}} \leq q \leq \frac{J^3}{3 \sqrt{3}} .
\end{equation}
This equation gives the upper boundary for the quantized values of
integral of motion shown in Fig.\ \ref{AbsQ}. This picture demonstrates
as well that there exist two possibilities of parametrization of the
spectrum by different sets of trajectories.

The trajectories following the $q = 0$ behave at $J \to \infty$ as
$q \propto J^2$. Introducing the convention $q^\star
\equiv q/\eta^2$ and the continuum variable $\tau = j/J$ the recursion
relation can be reduced in leading-$J$ approximation to a simple
first order differential equation the solution to which reads
\begin{equation}
\label{WKBfuncLow}
{\mit\Upsilon}_j = J i^j \phi (\tau) ,
\qquad
\phi (\tau)
= e^{- i \varphi}
\left| \Gamma \left( 3/2 + i q^\star \right) \right|^{- 2}
\left[ \tau^2 (1 - \tau^2) \right]^{1/2}
\left( \frac{1 - \tau^2}{\tau^2} \right)^{i q^\star} .
\end{equation}
It is valid for whole interval of $j$ namely for $j \gg 1$, $J - j \gg 1$
except of the vicinities of the reflection points $j_{\rm end} \sim 1, J$
where oscillating WKB solution is not applicable and we have to solve
Eq.\ (\ref{RecRel}) close to the end points exactly. We have for
$J \gg J - j \sim 1$
\begin{equation}
\label{LimitRecTwo}
2 q^\star {\mit\Upsilon}_j
= (J - j + 2) {\mit\Upsilon}_{j + 1}
+ (J - j + 1) {\mit\Upsilon}_{j - 1} ,
\end{equation}
while for $J \gg j \sim 1$
\begin{equation}
\label{LimitRecOne}
2 q^\star {\mit\Upsilon}_j
= \frac{(j + 1)(j + 3)}{2j + 5} {\mit\Upsilon}_{j + 1}
+ \frac{(j + 2)(j + 4)}{2j + 5} {\mit\Upsilon}_{j - 1} .
\end{equation}

To solve the first limiting recursion relation we make the following
ansatz
\begin{equation}
{\mit\Upsilon}_j = i^{J - j} \int\limits_{0}^{1} d \lambda
\lambda^{i q^\star - 1} (1 - \lambda)^{- i q^\star - 1}
\upsilon_{J - j} (\lambda) .
\end{equation}
Substituting it into Eq.\ (\ref{LimitRecTwo}) we get the equation
\begin{equation}
2 \lambda (1 - \lambda) \frac{d}{d \lambda} \upsilon_\ell (\lambda)
= (\ell + 1) \upsilon_{\ell + 1} (\lambda)
- (\ell + 2) \upsilon_{\ell - 1} (\lambda) ,
\end{equation}
with the solution
\begin{equation}
\upsilon_\ell (\lambda)
= C [\lambda (1 - \lambda)]^{-1/2} (1 - 2 \lambda)^{\ell + 2} .
\end{equation}

The solution to the second recurrent relation can be found by exploiting
the cyclic symmetry of the problem and the known WKB solution
(\ref{WKBfuncLow}). Using the Racah coefficients, $W_{jk}$, which relate
the different three-particle bases (\ref{RacahRelation}), we can write
\begin{equation}
{\mit\Psi}_j = e^{i \varphi}
\sum_{k = 0}^{J} W_{jk} (J) {\mit\Psi}_k .
\end{equation}
From the recursion relation which is obeyed by the coefficients $W$ (see
e.g.\ Refs. \cite{BraDerKorMan99,Bel99b}) we can obtain the formula for
the limit $\tau \equiv \frac{k}{J} = {\rm fixed}$, $j = {\rm fixed}$,
$J \to \infty$, namely,
\begin{equation}
\label{RacahMiddle}
W_{jk} = \frac{1}{\sqrt{J}} (- 1)^J w_j (\tau),
\quad\mbox{with}\quad
w_j (\tau) =
\sqrt{2}
\left[
\frac{\tau^5 (1 - \tau^2)^2 }{N_j}
\right]^{1/2}
C_j^{5/2} (2 \tau^2 - 1),
\end{equation}
with the normalization coefficient $N_j^{-1} = (3 \cdot 4)^2
(2 j + 5)/(j + 1)_4$. Then the solution for $j \sim 1$ region
is given by
\begin{equation}
{\mit\Upsilon}_j = i^j (- 1)^J e^{i \varphi}
\sqrt{2}
\left[ \frac{(j + 1)_4}{2 j + 5}\right]^{1/2}
\int\limits_{0}^{1} d \tau \,
\frac{\phi (\tau) w_j (\tau)}{[ \tau^3 (1 - \tau^2)^2 ]^{1/2}}.
\end{equation}
A simple calculation gives us finally the result
\begin{equation}
{\mit\Upsilon}_j = (- 1)^{J - j}
\frac{(j + 1)_4}{4} \,
p_j \left( q^\star \left| \frac{3}{2}, \frac{3}{2} \right)\right.
\end{equation}
in terms of Askey-Wilson polynomials\footnote{See also Refs.\
\cite{FadKor95,Kor95a} in connection to solution of the two-reggeon
Baxter equation.} \cite{AskWil85}
\begin{equation}
p_j \left( x | \alpha, \beta \right)
= i^j {_3F_2} \left( \left.
{ - j ,\ j + 2 \alpha + 2 \beta - 1 ,\ \alpha - i x
\atop
\alpha + \beta ,\ 2 \alpha } \right| 1 \right) ,
\end{equation}
which are orthogonal on the interval $- \infty < x < \infty$
\begin{equation}
\int\limits_{- \infty}^{\infty} dx \,
p_j \left( x | \alpha, \beta \right)
p_k \left( x | \alpha, \beta \right)
| \Gamma (\alpha + i x) \Gamma (\beta + i x) |^2
= 0, \qquad j \not = k .
\end{equation}

Matching the WKB with the end-point solutions in the region of
their overlap we get the quantization conditions for the charge
\begin{equation}
\label{QuantizedQbottom}
q^\star \ln \eta
= \arg \, \Gamma \left( \frac{3}{2} + i q^\star \right)
+ \frac{\pi}{6} \left( 2 m + \sigma_J \right) ,
\end{equation}
where $\sigma_J = [ 1 - (- 1)^J ]/2$. The first iteration at large
$\eta$ is
\begin{equation}
\label{LowChargeApp}
q^\star = \frac{\pi}{6} \frac{ 2 m + \sigma_J}{\ln \eta - \psi (3/2)}
\end{equation}
with accuracy $\cO (\ln^{- 3} \eta)$.

Let us turn to the second way of describing the spectrum of $q$. The
maximum of the eigenfunctions describing the trajectories which behave as
$q \to J^3/\sqrt{27}$ is achieved for $j_{\rm max} = \frac{J}{\sqrt{3}}$
and in the vicinity of this point the system behaves as a classical one.
To find the quasiclassical corrections we expand the recursion relation
around this point $j = \frac{1}{\sqrt{3}} \left( J + \lambda \sqrt{J}
\right)$, and look for the solution to Eq.\ (\ref{RecRel}) in the form
of the series w.r.t.\ the inverse powers of $J$
\begin{equation}
\label{WKBexp}
{\mit\Phi} (\lambda) = \sum_{\ell = 0}^{\infty}
{\mit\Phi}_{(\ell)} (\lambda) J^{- \ell/2} ,
\qquad\mbox{and}\qquad
q(J, n) = \frac{J^3}{\sqrt{3}}
\sum_{\ell = 0}^{\infty} q^{(\ell)}(n) J^{- \ell},
\end{equation}
where ${\mit\Upsilon}_j \equiv {\mit\Phi} (\lambda)$.
This leads to the infinite sequence of differential equations
\begin{eqnarray}
\label{FirstEq}
&&\cD_{(1)} {\mit\Phi}_{(0)} (\lambda) = 0, \nonumber\\
\label{SecondEq}
&&\cD_{(1)} {\mit\Phi}_{(1)} (\lambda)
+ \cD_{(2)} {\mit\Phi}_{(0)} (\lambda) = 0, \nonumber\\
&&\cD_{(1)} {\mit\Phi}_{(2)} (\lambda)
+ \cD_{(2)} {\mit\Phi}_{(1)} (\lambda)
+ \cD_{(3)} {\mit\Phi}_{(0)} (\lambda) = 0, \\
&&\dots , \nonumber
\end{eqnarray}
supplied with the boundary conditions ${\mit\Phi} (\pm\infty) = 0$
and the differential operators given by the formulae
\begin{eqnarray}
\cD_{(1)} \!\!\!&=&\!\!\! \frac{d^2}{d \lambda^2}
+ \left( 8 - 2 q^{(1)} - \lambda^2 \right) ,
\nonumber\\
\cD_{(2)} \!\!\!&=&\!\!\! - \frac{d}{d \lambda}
+ \lambda \left( 8 - 5 \sqrt{3} - \frac{1}{3} \lambda^2 \right) ,
\nonumber\\
\cD_{(3)} \!\!\!&=&\!\!\! \frac{1}{4} \frac{d^4}{d \lambda^4}
+ \frac{3}{2} \left( 8 - \lambda^2 \right) \frac{d^2}{d \lambda^2}
+ 4 \lambda \frac{d}{d \lambda}
- \frac{1}{2}
\left( 11 - 40 \sqrt{3} + 4 q^{(2)} + 5 \sqrt{3} \lambda^2 \right) ,
\nonumber\\
\dots&&
\nonumber
\end{eqnarray}
The solutions ${\mit\Phi}_{(\ell)} (\lambda) = \phi_{(\ell)} (\lambda)
\exp(- \lambda^2/2)$ to these equations are expressed by a linear
combination of the Hermite polynomials
\begin{eqnarray}
\label{Hermite0}
\phi_{(0)} (\lambda) &=& H_n \left( \lambda \right) , \\
\label{Hermite1}
\phi_{(1)} (\lambda)
&=& \frac{1}{18} n(n - 1)(n - 2) H_{n - 3} (\lambda)
+ \frac{n}{4} (n - 14 + 10 \sqrt{3}) H_{n - 1} (\lambda) \nonumber\\
&-& \frac{1}{8} (n - 17 + 10 \sqrt{3}) H_{n + 1} (\lambda)
- \frac{1}{144} H_{n + 3} (\lambda), \\
\dots\quad &&\nonumber
\end{eqnarray}
where $n =0,1,\dots$ and gives the number of nodes of the solutions in
the classically allowed region. Deriving these equations we have
implicitly assumed $n \ll J$ in order the expansion (\ref{WKBexp})
for the charge to be meaningful.

\begin{figure}[t]
\unitlength1mm
\begin{center}
\vspace{-1cm}
\hspace{0cm}
\begin{picture}(100,155)(0,0)
\put(0,60){\insertfig{9}{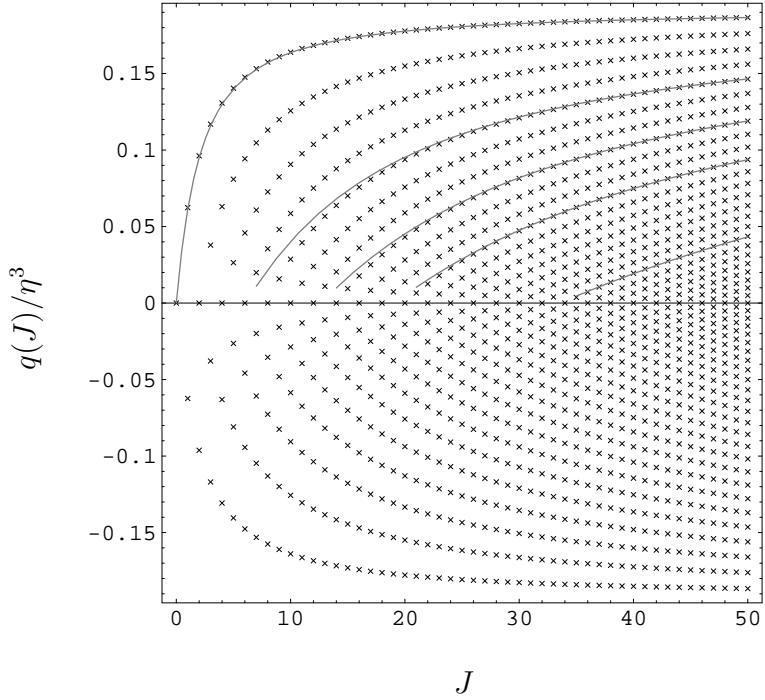}}
\put(49,55){$J$}
\put(-10,98){\rotate{$q (J)/\eta^3$}}
\end{picture}
\end{center}
\vspace{-6cm}
\caption{\label{Qcharge} The numerical quantized values of the charge
(crosses) and WKB approximation (\ref{WKBexp},\ref{QfewWKBcorr}).}
\end{figure}

The quantization for the charge $q$ immediately follows from
boundary conditions. In this way we derive a few next-to-leading
corrections to the charge
\begin{eqnarray}
\label{QfewWKBcorr}
q^{(0)} (n)
&=& \frac{1}{3} , \nonumber\\
q^{(1)} (n)
&=& \frac{7}{2} - n , \nonumber\\
q^{(2)} (n)
&=& \frac{85}{9} - \frac{22}{3} n + \frac{2}{3} n^2 ,\\
\dots ,\ \ \ && \nonumber
\end{eqnarray}
which compare quite well with the eigenvalues deduced from the numerical 
diagonalization of Eq.\ (\ref{RecRel}), see Fig.\ \ref{Qcharge}.

\subsection{Eigenvalues of Hamiltonian $\cH_0$.}

Having found the expansion coefficients ${\mit\Upsilon}_j$, which
completely specify the eigenfunction ${\mit\Psi}_0$, we can express
the three-particle energy $\cE_0$ in their terms by exploiting the
permutation symmetry of the system
\begin{equation}
\label{EnegyLinear}
\cE_0 (J, q) = 2 {\rm Re}\,
\frac{\sum_{j = 0}^{J} (- i)^j (2 j + 5) \epsilon (j) {\mit\Upsilon}_j}{
\sum_{j = 0}^{J} (- i)^j (2 j + 5) {\mit\Upsilon}_j} + 3 ,
\end{equation}
with the two-particle energy
\begin{equation}
\epsilon (j) = 2 \psi (j + 3) - 2 \psi (1) .
\end{equation}
Conventionally, taking the average of the Hamiltonian $\cH_0$ w.r.t.\ 
the eigenfunctions ${\mit\Psi}_0$ we find an equivalent expression
for the energy
\begin{equation}
\label{EnegySquare}
\cE_0 (J, q) = 3
\frac{\sum_{j = 0}^{J} \epsilon (j) |{\mit\Psi}_j|^2}{
\sum_{j = 0}^{J} |{\mit\Psi}_j|^2} .
\end{equation}
These results provide the exact energy of the three-gluon system with
the same helicity of all particles.

\begin{figure}[t]
\unitlength1mm
\begin{center}
\vspace{-1cm}
\hspace{0cm}
\begin{picture}(100,155)(0,0)
\put(0,60){\insertfig{9}{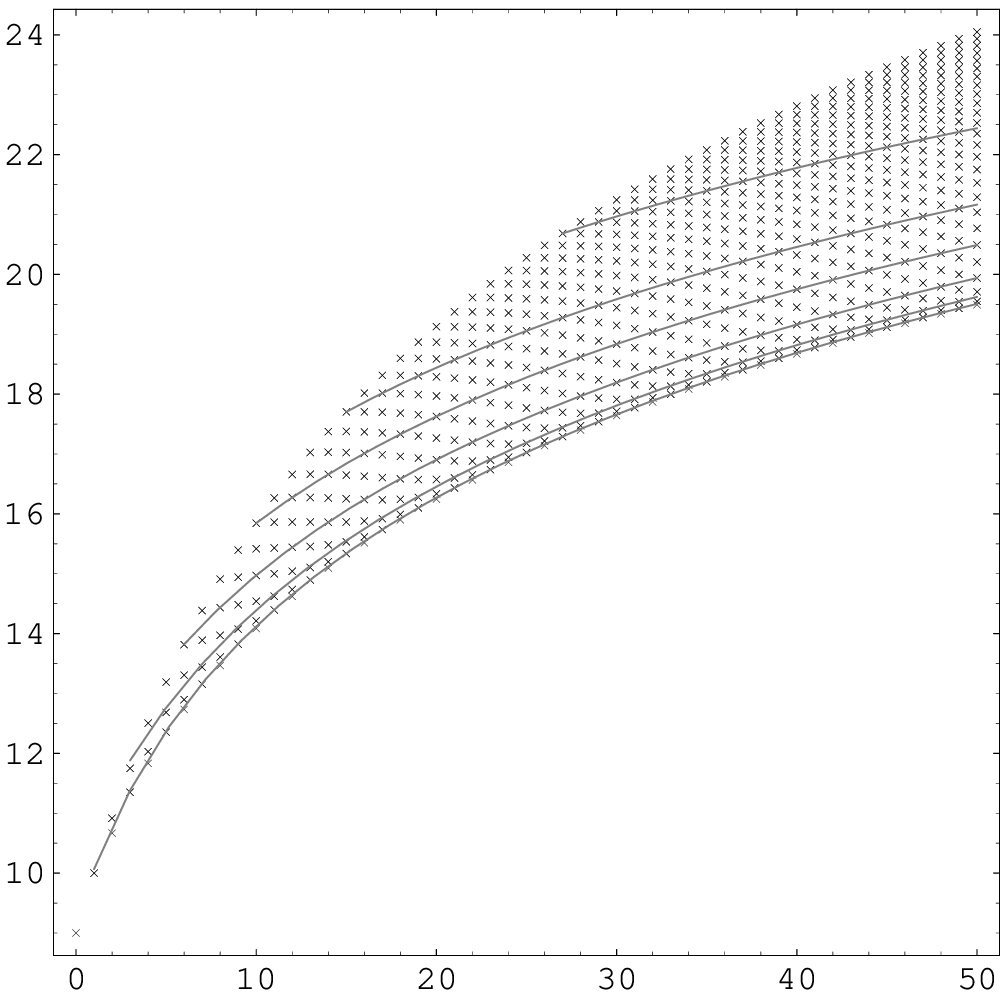}}
\put(46,55){$J$}
\put(-10,100){\rotate{$\cE_0 (J)$}}
\end{picture}
\end{center}
\vspace{-6cm}
\caption{\label{BotomXXXenergy} The numerical eigenvalues of the
Hamiltonian $\cH_0$ versus the analytical energy given in Eq.\
(\ref{EnergyBottomExpand}) for $m = 0, 1$ (odd $J$) and Eq.\
(\ref{EnergyBottom}) for higher trajectories $m = 3, 5$
(even $J$), $m = 7, 13$ (odd $J$).}
\end{figure}

Substituting the eigenfunction (\ref{EigenfunQzero}) corresponding to
the special case when $q = 0$ we easily obtain the lowest energy
trajectory for even conformal spin $J$
\begin{equation}
\label{Qoenergy}
\cE_0 (J,0) = 2 \psi \left( \frac{J}{2} + 3 \right)
+ 2 \psi \left( \frac{J}{2} + 2 \right) - 4 \psi (1) + 4 .
\end{equation}
Next, using the WKB solutions found in the preceding section we can
get the explicit result for the set of trajectories describing the
spectrum from below using Eq.\ (\ref{WKBfuncLow}) as
\begin{equation}
\label{EnergyBottom}
\cE_0 (J, q) = 4 \ln \eta - 6 \psi (1)
+ 2 {\rm Re}\, \psi \left( \frac{3}{2} - i q^\star \right) ,
\end{equation}
where $q^\star \equiv q/\eta^2$ and is valid with $\cO (\eta^{-1})$
accuracy. For the few lowest trajectories following the $q = 0$ one
we can write from here the approximate formula
\begin{equation}
\label{EnergyBottomExpand}
\cE_0 (J, q) = 4 \ln \frac{\eta}{2} - 4 \psi (1) + 4
- \psi'' \left( 3/2 \right) \frac{\pi^2}{36}
\left(
\frac{2 m + \sigma_J}{\ln \eta - \psi (3/2)}
\right)^2 ,
\end{equation}
where we have used the explicit form of the quantized $q$ from Eq.\
(\ref{LowChargeApp}). For comparison with the numerical diagonalization
see Fig.\ \ref{BotomXXXenergy}.

For the trajectories starting from the top of the energy spectrum we
get using Eqs.\ (\ref{Hermite0},\ref{Hermite1}) and (\ref{EnegySquare})
\begin{equation}
\label{WKBtopRed}
\cE (J, q) = 2 \ln q - 6 \psi (1) + \cO (J^{- 2}) ,
\end{equation}
where we have to substitute the explicit WKB expansion (\ref{QfewWKBcorr})
for the integral of motion. Evaluation of the higher WKB corrections to
this equation is extremely complicated task because of the lengthy
form of the WKB functions $\phi_{(\ell)}$ for $\ell \geq 2$ so that
even $\cO (J^{- 2})$ correction to energy becomes intractable.

\subsection{Bethe ansatz and Baxter equation.}

As we have seen above the evaluation of the higher WKB correction
to the energy (\ref{WKBtopRed}) is a rather non-trivial task within the
formalism used above as it requires cumbersome calculations due to a
badly organized expansion in $J$: the actual expansion parameter for the
eigenfunctions was $\sqrt{J}$. In this section we will address this
issue starting from the Baxter equation for the $Q$-operator. For
the polynomial solutions the latter is equivalent to the usual algebraic
Bethe ansatz \cite{SklTakFad79,Fad80,FadTak79,KorBogIze93,Fad95}.

To construct the diagonalized auxiliary transfer matrix $t_{\frac{1}{2}}
(\lambda)$ one defines a local vacuum state on each site $| \omega_\ell
\rangle$ so that\footnote{Note that we can equally treat $\hat J^-_\ell$
as the creation and $\hat J^+_\ell$ as the annihilation operators,
respectively, as is done conventionally \cite{Fad95}. This corresponds to
the second choice of the pseudovacuum. In this case the local vacuum is
defined as $| \omega_\ell \rangle = \theta_\ell^{- \nu - 1}$ and the
eigenvalue of $\hat J^3_\ell$ just changes the sign as compared to the
Eqs.\ (\ref{LocalVacuum}): $\hat J^3_\ell | \omega_\ell \rangle = -
\frac{\nu + 1}{2} | \omega_\ell \rangle$. In this case the Bethe states
are $| {\mit\Psi}_J \rangle = \prod_{\ell = 1}^{J} B (\lambda_\ell) |
{\mit\Omega} \rangle$ and the eigenvalues of the auxiliary transfer
matrix as well as the Bethe Ansatz equation remains intact.}
\begin{equation}
\label{LocalVacuum}
\hat J^-_\ell | \omega_\ell \rangle
= 0, \qquad
\hat J^3_\ell | \omega_\ell \rangle
= \frac{\nu + 1}{2} | \omega_\ell \rangle ,
\end{equation}
where from the definitions of the $su(1,1)$ generators we conclude that
$| \omega_\ell \rangle =\, $const but it is not annihilated by the
generator of the conformal spin $\mbox{\boldmath$\hat J$}$ and thus is
a legitimate choice. Then the vacuum of the chain is $| {\mit\Omega}
\rangle = \OO_{\ell = 1}^3 | \omega_\ell \rangle$ and the auxiliary
monodromy matrix acts as
\begin{equation}
T_{\frac{1}{2}} (\lambda) | {\mit\Omega} \rangle
\equiv
\left(
\begin{array}{cc}
A (\lambda) & B (\lambda) \\
C (\lambda) & D (\lambda)
\end{array}
\right)
| {\mit\Omega} \rangle
=
\left(
\begin{array}{cc}
\left( \lambda + \frac{\nu + 1}{2} \right)^3
&
0
\\
\star
&
\left( \lambda - \frac{\nu + 1}{2} \right)^3
\end{array}
\right)
| {\mit\Omega} \rangle ,
\end{equation}
where the $\star$ stands for a term whose explicit form does not matter
for us. Hence, as we see the element $B (\lambda)$ is the annihilation
operator while $C (\lambda)$ can treated as the creation one. Therefore, 
we can construct a special set of the states of the chain by acting with
$C (\lambda)$ on the vacuum $| {\mit\Omega} \rangle$, the so-called Bethe
states
\begin{equation}
| {\mit\Psi}_J \rangle
= \prod_{\ell = 1}^{J} C (\lambda_\ell) | {\mit\Omega} \rangle ,
\end{equation}
whose spin is $\sum_{\ell = 1}^{3} \hat J^3_\ell | {\mit\Psi}_J \rangle
= (J + \frac{3}{2}(\nu + 1) ) | {\mit\Psi}_J \rangle$ \cite{Fad95}. Making
use of the exchange relations stemming from the $RTT$ fundamental
commutation relation one can easily deduce that these states are the
eigenvectors of the auxiliary transfer matrix $t_{\frac{1}{2}} (\lambda)
= A (\lambda) + D (\lambda)$ with the eigenvalues
\cite{SklTakFad79,FadTak79,KorBogIze93,Fad95}
\begin{equation}
\label{EigenvaluesTransfer}
t_{\frac{1}{2}} (\lambda)
= \left( \lambda + \frac{\nu + 1}{2} \right)^3
\prod_{\ell = 1}^{J}
\frac{\lambda - \lambda_\ell + 1}{\lambda - \lambda_\ell}
+
\left( \lambda - \frac{\nu + 1}{2} \right)^3
\prod_{\ell = 1}^{J}
\frac{\lambda - \lambda_\ell - 1}{\lambda - \lambda_\ell} ,
\end{equation}
provided the Bethe roots $\{ \lambda_\ell | \ell = 1,\dots, J\}$ satisfy
the Bethe Ansatz equation
\begin{equation}
\left( \lambda_m + \frac{\nu + 1}{2} \right)^3
\prod_{{\ell = 1 \atop \ell \not = m}}^{J}
\left( \lambda_m - \lambda_\ell + 1 \right)
=
\left( \lambda_m - \frac{\nu + 1}{2} \right)^3
\prod_{{\ell = 1 \atop \ell \not = m}}^{J}
\left( \lambda_m - \lambda_\ell - 1 \right) .
\end{equation}
On this condition the Bethe states become degree-$J$ homogeneous
translation invariant polynomials in $\theta_\ell$.

By virtue of Eq.\ (\ref{Commutativity}) one can assume that the Bethe
states are also the eigenstates of the fundamental transfer matrix. It
is really the case and for the Hamiltonian of the lattice we have the
expression in terms of Bethe roots \cite{TarTakFad83,Fad95}
\begin{equation}
\cE_0 = \sum_{\ell = 1}^{J} \frac{d}{d \lambda_\ell}
\ln \frac{\lambda_\ell + \frac{\nu + 1}{2}}{
\lambda_\ell - \frac{\nu + 1}{2}}
+ 3 \left. \frac{d}{d \lambda} \right|_{\lambda = 0} f (\nu, \lambda) .
\end{equation}

Introducing finally the function
\begin{equation}
\label{BaxterPolynom}
Q (\lambda) = \prod_{\ell = 1}^{J} \left( \lambda - \lambda_\ell \right),
\end{equation}
we have from Eq.\ (\ref{EigenvaluesTransfer}) the Baxter equation
\cite{Bax82,Skl85,Skl92}
\begin{equation}
\label{BaxterEq}
t_{\frac{1}{2}} (\lambda) Q (\lambda)
=
\left( \lambda + \frac{\nu + 1}{2} \right)^3
Q (\lambda + 1)
+
\left( \lambda - \frac{\nu + 1}{2} \right)^3
Q (\lambda - 1)
\end{equation}
for the eigenvalues of the $Q$-operator where the $t_{\frac{1}{2}}
(\lambda)$ stands for the eigenvalues of the transfer matrix, rather
then the operator, and reads
\begin{equation}
t_{\frac{1}{2}} (\lambda)
= 2 \lambda^3
+ \left( \eta^2 - 3 {\mbox{\boldmath$\hat \jmath$}}^2 \right) \lambda
+ i q .
\end{equation}
The usefulness of the Baxter operator is that the energy of the system
can be rewritten concisely in its terms making use of the known Bethe
ansatz representation \cite{FadKor95}
\begin{equation}
\cE_0 = \frac{Q' \left( \frac{\nu + 1}{2} \right)}{
Q \left( \frac{\nu + 1}{2} \right)}
-
\frac{Q' \left( - \frac{\nu + 1}{2} \right)}{
Q \left( - \frac{\nu + 1}{2} \right)}
+ 6 \psi (\nu + 1) - 6 \psi (1) ,
\end{equation}
provided $f (\nu, \lambda)$ is given by the expression after Eq.\
(\ref{TwoSiteHamilton}). Thus, the knowledge of the Baxter operator
as a function of the conserved charges allows to find immediately
the energy as a function of the latter. The Eq.\ (\ref{BaxterEq})
will be the main object of our analysis in this section.

\subsection{Quasiclassical expansion.}

Let us analyze the Baxter equation for large values of the total
conformal spin, $J$, of the three-particle system as was done before
when we have dealt with the recursion relation (\ref{RecRel}). In these
case the system behaves as a quasiclassical one, and we can apply
appropriate methods used previously for the analysis of the Toda
\cite{PasGau92} and $s = - 1$ Heisenberg spin chain \cite{Kor95b}.

To get rid of the large factor $J$ in the transfer matrix let us
rescale the spectral parameter as $\lambda \to J \lambda$ and
introduce instead of $Q$ and $t_{\frac{1}{2}}$ the functions
\begin{equation}
\label{RescaledFunct}
\xi (\lambda) = Q (J \lambda),
\qquad
\cT (\lambda) = (J \lambda)^{- 3} t_{\frac{1}{2}} (J \lambda).
\end{equation}
Then Eq.\ (\ref{BaxterEq}) takes the form
\begin{equation}
\label{RedBaxterEq}
\left( \lambda + \frac{\nu + 1}{2} J^{- 1}\right)^3
\xi (\lambda + J^{- 1})
+
\left( \lambda - \frac{\nu + 1}{2} J^{- 1}\right)^3
\xi (\lambda - J^{- 1})
= \lambda^3 \cT (\lambda) \xi (\lambda) ,
\end{equation}
where $\cT (i \lambda)$ is a real function. Moreover we introduce
conventionally the WKB function as
\begin{equation}
\xi (\lambda) = e^{J S (\lambda)} ,
\end{equation}
so that the energy is expressed now as
\begin{equation}
\label{EnergyFromS}
\cE_0
= S' \left( \frac{\nu + 1}{2} J^{- 1} \right)
- S' \left( - \frac{\nu + 1}{2} J^{- 1} \right)
+ 6 \psi (\nu + 1) - 6 \psi (1) ,
\end{equation}
and, thus, for the limit $J \to \infty$ we can primarily be interested
in the small $\lambda$ asymptotics of Eq.\ (\ref{RedBaxterEq}). We look
the solution to this equation in terms of series w.r.t.\ inverse powers
of the conformal spin
\begin{equation}
\label{WKBexpansionST}
S (\lambda)
= \sum_{\ell = 0}^{\infty} S_{(\ell)} (\lambda) J^{- \ell}
\qquad\mbox{with}\qquad
\cT (\lambda)
= \sum_{\ell = 0}^{\infty} \cT_{(\ell)} (\lambda) J^{- \ell} ,
\end{equation}
where
\begin{eqnarray}
\label{DefRescTransMat}
\cT_{(0)} (\lambda) \!\!\!&=&\!\!\! 2 + \lambda^{- 2}
+ \frac{i}{\sqrt{3}} q^{(0)} \lambda^{-3}, \qquad
\cT_{(1)} (\lambda) = 8 \lambda^{- 2}
+ \frac{i}{\sqrt{3}} q^{(1)} \lambda^{-3}, \nonumber\\
\cT_{(2)} (\lambda) \!\!\!&=&\!\!\! \frac{27}{2} \lambda^{- 2}
+ \frac{i}{\sqrt{3}} q^{(2)} \lambda^{-3}, \qquad\ \
\cT_{(\ell)} (\lambda) =
\frac{i}{\sqrt{3}} q^{(\ell)} \lambda^{-3},
\quad\mbox{for}\quad \ell \geq 3 .
\end{eqnarray}
In leading order approximation we have the equation for $S'_{(0)}$
\begin{equation}
2 \cosh S'_{(0)} (\lambda) = \cT_{(0)} (\lambda) ,
\end{equation}
with the solution
\begin{equation}
\label{LOsolution}
S'_{(0)} (\lambda)
= 2 \ln \left\{
\sqrt{\frac{1}{4} \cT_{(0)} (\lambda) + \frac{1}{2} }
+ \sqrt{\frac{1}{4} \cT_{(0)} (\lambda) - \frac{1}{2} }
\right\} ,
\end{equation}
which has a correct asymptotics for $\lambda \to \infty$ since
$S (\lambda) \stackrel{\lambda \to \infty}{\sim} \ln \lambda$
as follows from Eq.\ (\ref{BaxterPolynom}) because $Q (\lambda)
\stackrel{\lambda \to \infty}{\sim} \lambda^J$. In the small
$\lambda$ region we get from Eq.\ (\ref{LOsolution})
\begin{equation}
\label{LOsolExpand}
S'_{(0)} (\lambda \to \pm 0)
= \pm \ln \frac{i}{\sqrt{3}} q^{(0)} \lambda^{- 3} .
\end{equation}
From here we can immediately find that for large conformal spin
$\cE_0 (J \to \infty) = 6 \ln J$ and coincides with the result
(\ref{WKBtopRed}). Unfortunately, due to the fact that for the
calculation of the energy from Eq.\ (\ref{EnergyFromS}) the function
$S$ enters in the point $\sim J^{- 1}$ in order to estimate a
non-leading correction we have to solve an infinite series of
differential equations stemming from Eq.\ (\ref{RedBaxterEq}). This
is, of course, not feasible. Therefore, we have to perform an effective
resummation of these series \cite{Kor95b}.

To do this we notice that from Eq.\ (\ref{LOsolExpand}) is follows
that for ${\rm Re} \lambda > 0$ we have $\exp\left( S'_{(0)}
(\lambda)\right) \gg \exp\left( - S'_{(0)} (\lambda)\right)$ and one
can easily convince oneself \cite{Kor95b} that up to $\cO (\lambda^6)$
the small $\lambda$ expansion of Eq.\ (\ref{RedBaxterEq}), i.e.\ Eqs.\
(\ref{WKBequationS}), can be substituted by the simplified equation
\begin{equation}
\frac{\xi \left( \lambda + J^{-1} \right)}{\xi (\lambda)}
= \lambda^3 \left( \lambda + \frac{\nu + 1}{2} J^{-1} \right)^{-3}
\cT (\lambda) .
\end{equation}
Similarly for ${\rm Re} \lambda < 0$ we can omit the first term on
the l.h.s.\ of Eq.\ (\ref{RedBaxterEq}). Now these simplified equation
can be solved exactly as
\begin{equation}
S' (\lambda)
= 3 \psi (J \lambda)
+ 3 \psi \left( J \lambda + \frac{\nu + 1}{2} \right)
+ \frac{ J^{- 1}\partial }{ \exp ( J^{- 1} \partial ) - 1 }
\ln \cT (\lambda) ,
\quad\mbox{for}\quad {\rm Re} \lambda > 0 .
\end{equation}
And analogously for the ${\rm Re} \lambda < 0$ obtained from this
equation by $J \to - J$ and changing the sign of the third term.
One can readily recognize the differential operator as the generating
function $t/(e^t - 1) = \sum_{\ell = 0}^{\infty} B_\ell t^\ell/ \ell !$
for the Bernoulli numbers of the first order \cite{BatErd53}.

Finally, we have for the energy
\begin{equation}
\label{BaxterEnergy}
\cE_0 = 6 \psi \left( \frac{\nu + 1}{2} \right) - 6 \psi (1)
+ 2 {\rm Re} \left.
\frac{ J^{- 1}\partial }{ \exp ( J^{- 1} \partial ) - 1 }
\right|_{\lambda = \frac{\nu + 1}{2} J^{- 1}}
\ln \cT (\lambda) + \cO (J^{- 6}).
\end{equation}
Making use of the representation of the auxiliary transfer matrix
in terms of its zeros $r_\ell$
\begin{equation}
t_{\frac{1}{2}} (\lambda)
= 2 \prod_{\ell = 1}^{3} (\lambda - r_\ell) ,
\end{equation}
one can easily obtain a concise expression of the energy (cf. \cite{Kor95b})
\begin{equation}
\cE_0 = 2 \ln 2 + 2 {\rm Re} \sum_{\ell = 1}^{3}
\left( \psi \left( \frac{\nu + 1}{2} - r_\ell \right)
- \psi (1) \right) + \cO (J^{- 6}) .
\end{equation}

\subsection{Quantization of $q$ and energy.}

\begin{figure}[t]
\unitlength1mm
\begin{center}
\vspace{-1cm}
\hspace{0cm}
\begin{picture}(100,155)(0,0)
\put(0,60){\insertfig{9}{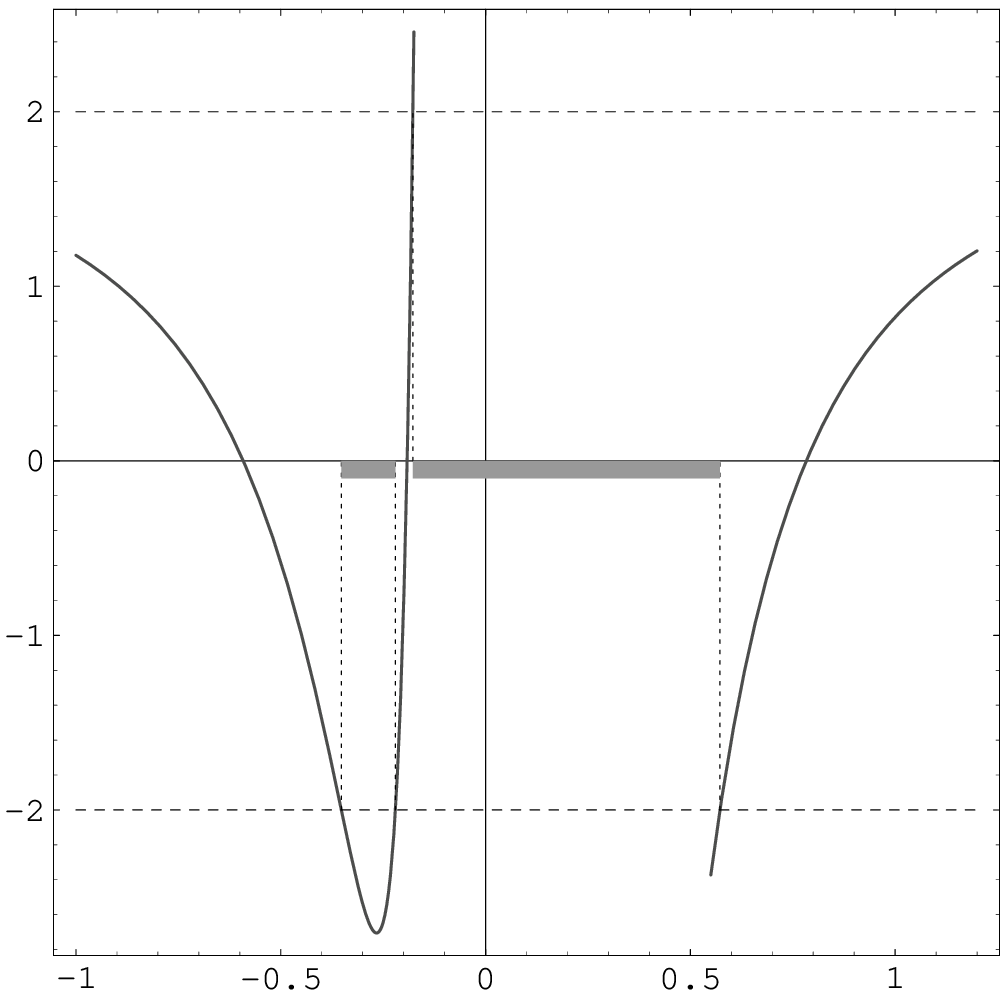}}
\put(29,110){${\scriptstyle\lambda_1}$}
\put(33,110){${\scriptstyle\lambda_2}$}
\put(38,110){${\scriptstyle\lambda_3}$}
\put(62,110){${\scriptstyle\lambda_4}$}
\put(46,55){$\lambda$}
\put(-10,100){\rotate{$\cT_{(0)} (i \lambda)$}}
\end{picture}
\end{center}
\vspace{-6cm}
\caption{\label{TransferMatrix} Typical form of the rescaled transfer
matrix (\ref{RescaledFunct}) for $0 < q^{(0)} < \frac{1}{3}$.}
\end{figure}

Expressing the energy directly in terms of the integrals of motion
we have to find their quantized values. Although it was done in
section \ref{RecursionRelationWKB} from the study of the recursion
relation we address this question here from the point of view of
the Baxter equation. From Eq.\ (\ref{BaxterPolynom}) we get for
$S' (\lambda)$
\begin{equation}
S' (\lambda) = \sum_{\ell = 1}^{J} ( J \lambda - \lambda_\ell )^{- 1},
\end{equation}
and thus
\begin{equation}
\oint\limits_{C} \frac{d \lambda}{2 \pi i} S' (\lambda) = 1 ,
\end{equation}
for a contour in the complex $\lambda$-plane which encircles all the
roots $\lambda_\ell/J$. From the leading WKB solution we conclude that
the zeros of $Q (\lambda)$ are accumulated on the interval
$|\cT (\lambda)| \geq 2$ on the imaginary axis. For the case at
hand we have two intervals of instability for the classical motion
\cite{FlaMcL76,PasGau92,Kor95b} and they are determined by the system
of cuts $[\lambda_1, \lambda_2] \cup [\lambda_3, \lambda_4]$. Therefore,
we have $n$ zeros on $[\lambda_1, \lambda_2]$ and $J - n$ on
$[\lambda_3, \lambda_4]$ and, thus,
\begin{equation}
\label{QuantCondanyJ}
\oint\limits_{C_1} \frac{d \lambda}{2 \pi i} S' (\lambda) = \frac{n}{J} ,
\end{equation}
with the contour $C_1$ which encircles the interval
$[\lambda_1, \lambda_2]$ (see Fig.\ \ref{TransferMatrix}). It is
obvious that $0 \leq n \leq J$. Let us assume for the following that
$n \sim \cO (J^0)$ and substitute the expansion (\ref{WKBexpansionST})
into quantization condition (\ref{QuantCondanyJ}) to get the sequence
\begin{equation}
\label{Siquantization}
\oint\limits_{C_1} \frac{d \lambda}{2 \pi i} S'_{(0)} (\lambda) = 0 ,
\qquad
\oint\limits_{C_1} \frac{d \lambda}{2 \pi i} S'_{(1)} (\lambda) = n ,
\qquad
\oint\limits_{C_1} \frac{d \lambda}{2 \pi i} S'_{(\ell)} (\lambda)
= 0 ,\quad\mbox{for}\quad \ell \geq 2 .
\end{equation}
The functions $S'_{(\ell)}$ can be easily deduced from the Baxter
equation (\ref{RedBaxterEq}) and read
\begin{eqnarray}
S'_{(1)} = \frac{\cT_{(1)}}{2 \sinh S'_{(0)}}
\!\!\!&-&\!\!\!
\frac{1}{2} S''_{(0)} \coth S'_{(0)} - \frac{9}{2} \lambda^{- 1} ,
\nonumber\\
S'_{(2)} = \frac{\cT_{(2)}}{2 \sinh S'_{(0)}}
\!\!\!&-&\!\!\! \left\{ \frac{1}{2} S''_{(1)}
+ \frac{1}{8} \left( S''_{(0)} \right)^2
+ \frac{1}{2} \left( S'_{(1)} \right)^2
+ \frac{9}{2} \lambda^{- 1} S'_{(1)}
+ \frac{27}{4} \lambda^{- 2}
\right\} \coth S'_{(0)} \nonumber\\
&-&\!\!\! \frac{1}{6} S'''_{(0)}
- \frac{1}{2} S'_{(1)} S''_{(0)}
- \frac{9}{4} \lambda^{- 1} S''_{(0)} ,
\label{WKBequationS} \\
\dots , \qquad\qquad\quad && \nonumber
\end{eqnarray}
and $S'_{(0)}$ has been found before in Eq.\ (\ref{LOsolution}). From
the quantization condition for $S'_{(0)}$ and Fig.\ \ref{TransferMatrix}
it is seen that the function $S'_{(0)}$ should not have singularities
on the interval $[\lambda_1, \lambda_2]$ which is the case provided
the latter shrinks into the point $\lambda_{\rm crit}$. These critical 
values are defined by the equations $\cT'_{(0)} (\lambda_{\rm crit}) = 0$ 
and $\cT_{(0)} (\lambda_{\rm crit}) = - 2$ with the solution
\begin{equation}
q_{\rm crit} \equiv q^{(0)} = \frac{1}{3}, \qquad
\lambda_{\rm crit} = - \frac{i}{2 \sqrt{3}} ,
\end{equation}
where we have restricted ourselves to $q^{(0)} > 0$.

To proceed further we consider an infinitesimal interval around
$\lambda_{\rm crit}$
\begin{equation}
\lambda_2 - \lambda_1 = i \varepsilon , \quad
\lambda_1 = \lambda_{\rm crit} - i \varepsilon/2 , \quad
\lambda_2 = \lambda_{\rm crit} + i \varepsilon/2 ,
\end{equation}
to ensure the presence of singularities of $S'_{(1)}$ on it and to fulfill 
the second quantization condition (\ref{Siquantization}). Here $\cT_{(0)} 
(\lambda_1) = \cT_{(0)} (\lambda_2) = -2$ as is seen from the Fig.\ 
\ref{TransferMatrix}. Expansion of the rescaled transfer matrix around 
the critical point $\lambda = \lambda_{\rm crit} + \frac{i}{2}
\varepsilon (2 x - 1)$ gives
\begin{equation}
\label{T0LO}
\cT_{(0)} = - 2 + 144 x (x - 1) \varepsilon^2 + \cO (\varepsilon^3) .
\end{equation}
Inserting it into Eqs.\ (\ref{LOsolution},\ref{WKBequationS}) we get
\begin{eqnarray}
\label{S01LO}
S'_{(0)} \!\!\!&=&\!\!\! i \pi + 12 i \sqrt{x (x - 1)} \varepsilon
+ \cO (\varepsilon^2) , \nonumber\\
S'_{(1)} \!\!\!&=&\!\!\! \frac{i}{24 \varepsilon}
\frac{\cT_{(1)} (\lambda_{\rm crit})}{\sqrt{x (x - 1)}}
+ \frac{i}{4 \varepsilon} \frac{2 x - 1}{x (x - 1)}
+ \cO (\varepsilon^0) .
\end{eqnarray}
Substituting the last equation into the quantization condition
$i \varepsilon \oint_{C_1} \frac{d x}{2 \pi i} S'_{(1)} (\lambda) = n$
we can deform the integration contour away from the cut $[\lambda_1,
\lambda_2]$ reverting its direction and take the residue at infinity
$x = \infty$ to get
\begin{equation}
- i \varepsilon \res_{x = \infty} \frac{i}{x \varepsilon}
\left\{ \frac{1}{24} \cT_{(1)} (\lambda_{\rm crit})
+ \frac{1}{2} \right\} = n .
\end{equation}
Using explicit form of $\cT_{(1)}$ from Eq.\ (\ref{DefRescTransMat})
one can immediately convince oneself that this is exactly $q^{(1)}$
in Eq.\ (\ref{QfewWKBcorr}).

Since the rescaled transfer matrix possesses the singularities only on the
interval $x \in [0,1]$ and infinity, we can always deform the integration
contour and evaluate the residue at $x = \infty$. Thus, for evaluation
of the quantized $q$ one has to keep only $(x \varepsilon)^{- 1}$-terms
in the expansions. This simplifies considerably the task of calculation
of the higher WKB corrections (and is easy to formalize) as compared to
the method based on the solution of the recursion relation in section
\ref{RecursionRelationWKB}. For instance, to find $q^{(2)}$ we have
just to keep more terms in the expansion in Eqs.\ (\ref{T0LO},\ref{S01LO}):
\begin{eqnarray}
\frac{1}{4} \cT_{(0)}
\!\!\!&=&\!\!\! - \frac{1}{2} + 36 (x \varepsilon)^2
+ 192 \sqrt{3} (x \varepsilon)^3
+ 2160 (x \varepsilon)^4 + \dots, \nonumber\\
\frac{1}{24} \cT_{(1)}
\!\!\!&=&\!\!\! \left( q^{(1)} - 4 \right)
+ 2 \sqrt{3} \left( 3 q^{(1)} - 8 \right) (x \varepsilon)
+ 72 \left( q^{(1)} - 2 \right) (x \varepsilon)^2
+ \dots,
\end{eqnarray}
which was done having in mind that we will be interested only in the
residue at $x = \infty$, $x \varepsilon = \,$fixed. And, thus, we have
\begin{eqnarray}
S'_{(0)} \!\!\!&=&\!\!\! i \left\{ \pi + 12 (x \varepsilon)
+ 32 \sqrt{3} (x \varepsilon)^2 + 304 (x \varepsilon)^3
+ \dots \right\} , \nonumber\\
S'_{(1)} \!\!\!&=&\!\!\! i \left\{
\frac{2 q^{(1)} - 7}{2 (x \varepsilon)}
+ \frac{10 q^{(1)} - 39}{\sqrt{3}}
+ \frac{2}{3} \left( 66 q^{(1)} - 239 \right) (x \varepsilon)
+ \dots \right\}
\end{eqnarray}
to the same accuracy. Therefore, we have finally the necessary terms
\begin{equation}
S'_{(2)} = \frac{i}{x \varepsilon}
\left\{
\frac{1}{24} \cT_{(2)} (\lambda_{\rm crit})
- \frac{2}{3} \left( q^{(1)} \right)^2 - \frac{8}{3} q^{(1)}
+ \frac{533}{36}
\right\} + \dots .
\end{equation}
From the requirement that this residue must equal zero we get the
result in Eq.\ (\ref{QfewWKBcorr}).

One can generalize in a straightforward way this consideration for
$q^{(\ell)}$ with $\ell \geq 3$, however, we will not pursue this goal
here since in the consequent consideration of the perturbed Hamiltonian
(\ref{PertHamilton}) we will be able to calculate the energy only up to
$\cO (J^{- 3})$ accuracy. The main motivation to study the Baxter
equation was to express the energy directly via the integral of
motion which allows to find the former in a very economic way.
Applying Eq.\ (\ref{BaxterEnergy}) we easily get
\begin{equation}
\cE_0 = 2 \ln \frac{q^{(0)} J^3}{\sqrt{3}}
+ \frac{2}{J} \frac{q^{(1)}}{q^{(0)}}
+ \frac{2}{J^2}
\left(
\frac{q^{(2)}}{q^{(0)}}
-
\frac{4 \left( q^{(1)}\right)^2 - 11}{8 \left( q^{(0)}\right)^2}
\right)
- 6 \psi (1)
+ \cO (J^{- 3}) ,
\end{equation}
where we have to plug in the explicit expressions for the
expansion coefficients of the charge (\ref{QfewWKBcorr}).

\subsection{Perturbed Hamiltonian.}

Let us turn to the perturbed Hamiltonian (\ref{PertHamilton}) whose
eigenvalues are responsible for the scale dependence of the three-gluon
correlator which contributes to the transverse spin structure function
$g_2$. Of course, the problem now is more complicated since the
interaction is not integrable. But it is easy to figure out that the
non-integrable addendum $\cV$ is small for a bulk of the spectrum and
this allows to consider it as a perturbation.

\begin{figure}[t]
\unitlength1mm
\begin{center}
\vspace{-1cm}
\hspace{0cm}
\begin{picture}(100,155)(0,0)
\put(0,60){\insertfig{9}{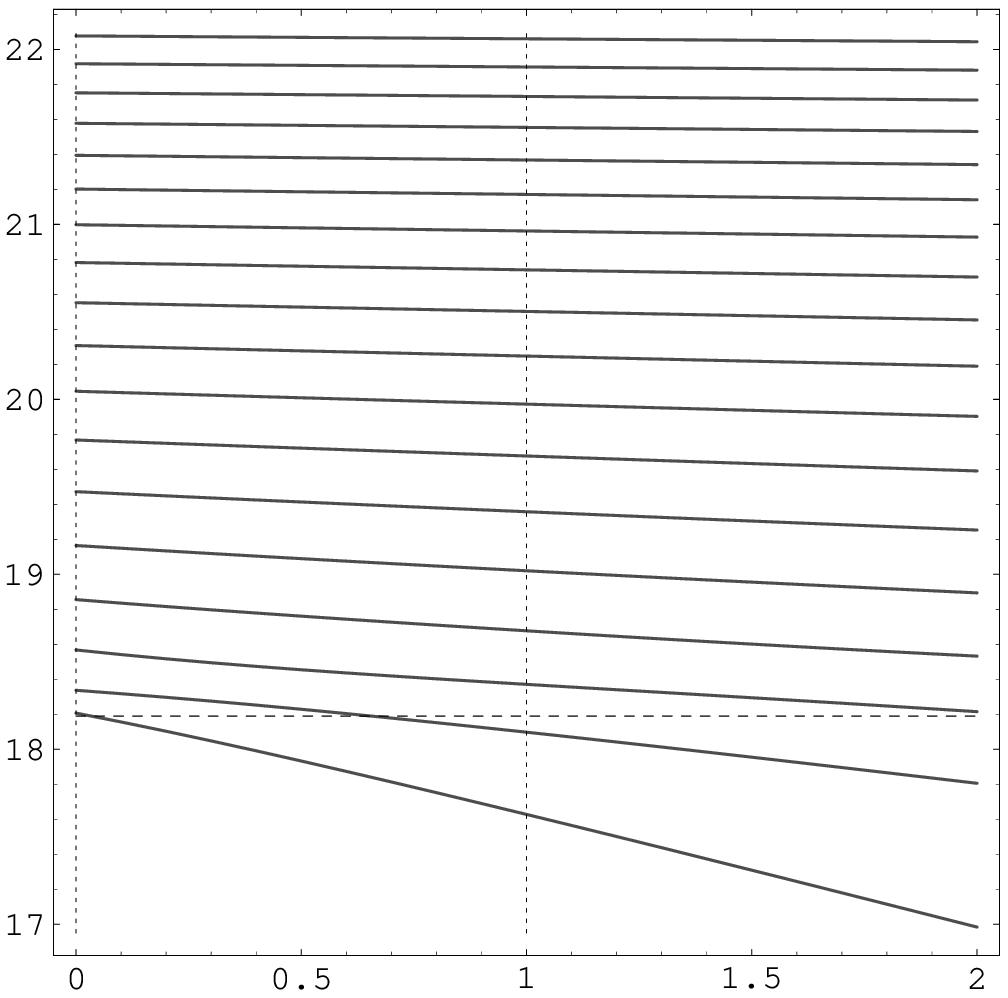}}
\put(15,70){\fbox{$\scriptstyle J = 35$}}
\put(46,55){$\alpha$}
\put(-10,106){\rotate{$\cE_\alpha$}}
\end{picture}
\end{center}
\vspace{-6cm}
\caption{\label{Flow} Flow of the eigenvalues of the $\cH_\alpha$ for
$J = 35$. The horizontal dashed line shows the position of the level
$\cE_0 (0)$.}
\end{figure}

To start with, let us introduce the more general Hamiltonian
\begin{equation}
\cH_\alpha = \cH + \alpha \cV
\end{equation}
specified by a coupling constant $\alpha$. Obviously $\cH_{\alpha = 0}
= \cH_0$ and $\cH_{\alpha = 1} = \cH$. Considering the flow of the
energy levels given in Fig.\ \ref{Flow} we conclude that the upper
part of the spectrum is not affected significantly by the perturbation
$\cV$ and it represents a small correction. At the same time it modifies
in a sizable way the level spacing for the lowest trajectories.

The present problem does not manifest the cyclic symmetry but
only w.r.t.\ the permutation of the first and third gluons.
Therefore, due to antisymmetry of the three-gluon operator we
will be interested in the eigenfunctions, ${\mit\Psi}^{(-)}$, of
the Hamiltonian (\ref{PertHamilton})
\begin{equation}
{\cH} {\mit\Psi}^{(-)} = \cE {\mit\Psi}^{(-)} ,
\end{equation}
which are odd functions w.r.t.\ the permutation of the end gluons
\begin{equation}
P_{13} {\mit\Psi}^{(-)} = - {\mit\Psi}^{(-)} .
\end{equation}
For the top of the spectrum we have $|\cV| \ll |\cH_0|$ for large $J$
and, therefore, as a first approximation we take
\begin{equation}
{\mit\Psi}^{(-)} = {\mit\Psi}^{(-)}_0 ,
\quad\mbox{with}\quad
{\mit\Psi}^{(-)}_0 ( q )
= \frac{1}{\sqrt{2}} \left\{
{\mit\Psi}_0 ( q )
-
{\mit\Psi}_0 ( - q )
\right\} .
\end{equation}
Then the matrix element of $\cV$ in this basis reads
\begin{equation}
\cV_{q',q} \equiv
\left[
\langle {\mit\Psi}^{(-)}_0 ( q' ) | {\mit\Psi}^{(-)}_0 ( q' ) \rangle
\langle {\mit\Psi}^{(-)}_0 ( q ) | {\mit\Psi}^{(-)}_0 ( q ) \rangle
\right]^{- 1/2}
\langle {\mit\Psi}^{(-)}_0 ( q' ) | \cV | {\mit\Psi}^{(-)}_0 ( q ) \rangle
\end{equation}
with the explicit eigenfunctions being plugged in there, it gives
\begin{eqnarray}
\cV_{q',q} \!\!\!&=&\!\!\! - 8 \left(
\sum_{j = 0}^{J}
| {\mit\Psi}_j (q') |^2
\sum_{j = 0}^{J}
| {\mit\Psi}_j (q) |^2
\right)^{- 1/2} \nonumber\\
&\times&\!\!\! \sum_{j = 0}^{J}
\frac{{\mit\Psi}_j^* (q') {\mit\Psi}_j (q)}{(j + 2)(j + 3)}
\left\{
\cos \left( \varphi(q') - \varphi(q) \right)
- (- 1)^j
\cos \left( \varphi(q') + \varphi(q) \right)
\right\} .
\end{eqnarray}
It defines the correction to the eigenfunctions
\begin{eqnarray*}
{\mit\Psi}^{(-)} (n)
=
{\mit\Psi}^{(-)}_0 (q)
+ \sum_{q' \neq q} \frac{\cV_{q',q}}{\cE_0 (q) - \cE_0 (q')}
{\mit\Psi}^{(-)}_0 (q') ,
\end{eqnarray*}
where $q = q (n)$ and $q' = q (n')$, and correction to the energy which
can be easily evaluated making use of Eqs.\ (\ref{Hermite0},\ref{Hermite1})
with the result
\begin{equation}
\cV_{q,q} = - \frac{24}{J^2} - \frac{48}{J^3} (2n - 3) + \cO (J^{- 4}) .
\end{equation}
Therefore, it affects the spacing of integrable Hamiltonian $\cH_0$ only 
at order $\cO(J^{- 2})$ while the net spacing is $\delta \cE_0 ( q \to J^3 
/ \sqrt{27}) \propto \cO (J^{-1})$ and the former is a small effect for 
large $J$.

As is seen from the picture of the flow of the energy eigenvalues of
the Hamiltonian $\cH (\alpha)$ the dependence on $\alpha$ is linear.
Thus, going to the limit $\alpha \to \infty$, we can consider the
integrable Hamiltonian as a correction on the background of $\cV$.
Calculating its matrix elements w.r.t.\ eigenfunctions of $\cV$ and
extrapolating the result to the point $\alpha = 1$ we can estimate
the spacing for the low trajectories of the $\cH$.

The eigenfunctions of the $\cV$ kernel which diagonalize it in
the limit of large $J$ are
\begin{equation}
{\mit\Psi}^{\cV (-)}_{J; m}
= \frac{1}{\sqrt{2}}
\left\{
\cP_{J; m} (\theta_1, \theta_2 | \theta_3)
-
\cP_{J; m} (\theta_3, \theta_2 | \theta_1)
\right\} ,
\end{equation}
with required antisymmetry w.r.t.\ the $P_{13}$ permutation. They are
normalized as $\langle {\mit\Psi}^{\cV (-)}_{J; m} |
{\mit\Psi}^{\cV (-)}_{J; m} \rangle = \delta_{m,m'} + \cO (J^{- 2})$.
We have the eigenvalues
\begin{equation}
\label{ScalarSpectrum}
\cV (m) \equiv
\langle {\mit\Psi}^{\cV (-)}_{J; m} |
\cV
| {\mit\Psi}^{\cV (-)}_{J; m} \rangle
/ \langle {\mit\Psi}^{\cV (-)}_{J; m} |
{\mit\Psi}^{\cV (-)}_{J; m} \rangle
= - \frac{4}{(m + 2)(m + 3)} + \cO (J^{- 2}) .
\end{equation}
The $\cO (J^{- 2})$ terms come form $\delta \cV (m) = \cV (m) +
\frac{4}{(m + 2)(m + 3)} = \frac{4 (- 1)^m W_{mm}}{(m + 2)(m + 3)}
- 4 \sum_{\ell = 0}^{J} \frac{W_{m \ell} W_{m \ell}}{(\ell + 2)(\ell + 3)}$
which can be estimated as follows. In the sum the main contributions
come for the regions $m, \ell \sim 1$ and $0 \leq \tau \leq 1$ with 
$\ell = \tau J$. For the latter we can replace the sum by the integral, 
$\sum \to J \int_0^1$, and use $W$ from Eq.\ (\ref{RacahMiddle}) so that 
one can immediately conclude that sum $\propto J^{- 2}$. For the small 
fixed $m$ and $\ell$, and asymptotical $J$, as can be deduced from the 
recursion relation obeyed by $W$, the Racah coefficients are given by
\begin{equation}
W_{jk} = \frac{(- 1)^J}{J^2} (2 k + 5) (j + 2)(j + 3) + \cO (J^{- 4}) .
\end{equation}
and, therefore, we have finally for the matrix element $\delta \cV (m)
\propto \cO (J^{- 2})$. This estimation is an example of the asymptotic
addition law for anomalous dimensions of scalar multi-particles composite
operators \cite{DerMan96}, according to which, for the present case, the
limiting point of the spectrum at large $J$ is given by the naive sum
of the anomalous dimension of a field and of a two-particle composite
operator. One can immediately recognize in Eq.\ (\ref{ScalarSpectrum})
the anomalous dimensions of the two-particle local composite operators
in $\phi_{(6)}^3$-theory.

Now the correction to the spectrum (\ref{ScalarSpectrum}) is given by
the $\alpha^{- 1} \langle {\mit\Psi}^{\cV (-)}_{J; m} | \cH_0
| {\mit\Psi}^{\cV (-)}_{J; m} \rangle$ and reads
\begin{equation}
\langle {\mit\Psi}^{\cV (-)}_{J; m} | \cH_0
| {\mit\Psi}^{\cV (-)}_{J; m} \rangle
/ \langle {\mit\Psi}^{\cV (-)}_{J; m} |
{\mit\Psi}^{\cV (-)}_{J; m} \rangle
= \epsilon (m) \left( 1 - (- 1)^m W_{mm} \right)
+ \sum_{\ell = 0}^{J} \left( 2 - (- 1)^\ell \right) \epsilon (\ell)
W_{m \ell} W_{m \ell}.
\end{equation}
Making use of the explicit asymptotics of the Racah coefficients we get
finally (up to $\cO (\eta^{- 1})$)
\begin{equation}
\cE_\alpha (m) = \alpha \cV (m)
+ 4 \ln \eta + 4 \psi (m + 3) + 2 \psi (m + 5)
- 2 \psi (2m + 5) - 2 \psi (2m + 6) - 6 \psi (1) .
\end{equation}
Since the dependence on $\alpha$ observed for the energy flow is
linear we can extrapolate this result to the point $\alpha = 1$
and estimate the spacing for the lowest trajectories of $\cE$.
For instance for $m = 0$ we get $\cE_1 (0) - \cE_0 (J, 0) = - 0.46$
which is rather close to the numerical value found from the explicit
diagonalization.

\begin{figure}[p]
\unitlength1mm
\begin{center}
\vspace{-1cm}
\hspace{0cm}
\begin{picture}(100,155)(0,0)
\put(0,60){\insertfig{9}{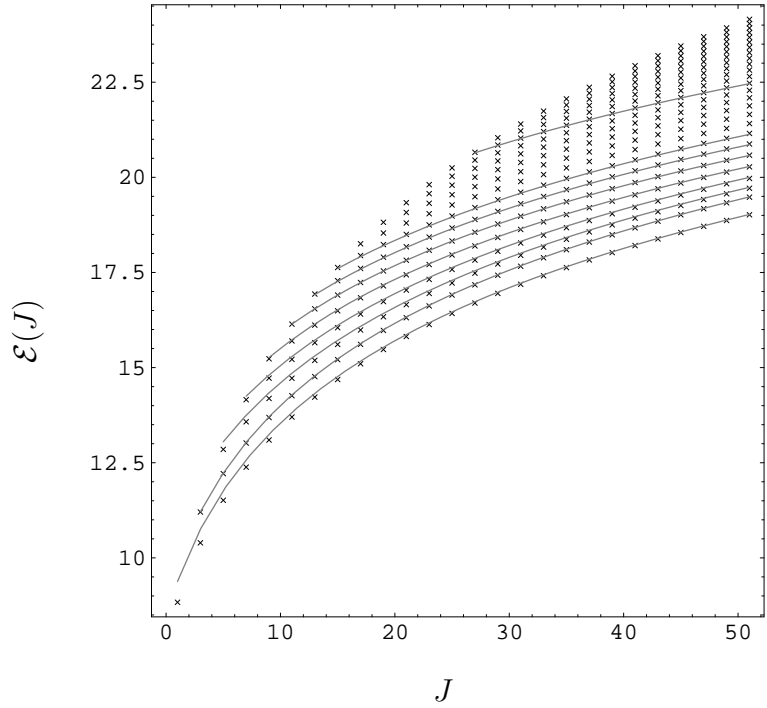}}
\put(46,55){$J$}
\put(-10,100){\rotate{$\cE (J)$}}
\end{picture}
\end{center}
\vspace{-6cm}
\caption{\label{G2energy} The exact energy eigenvalues versus the
approximate formulae (\ref{EnerQ2app}).}
\end{figure}

\begin{figure}[p]
\unitlength1mm
\begin{center}
\vspace{-1cm}
\hspace{0cm}
\begin{picture}(100,155)(0,0)
\put(0,60){\insertfig{9}{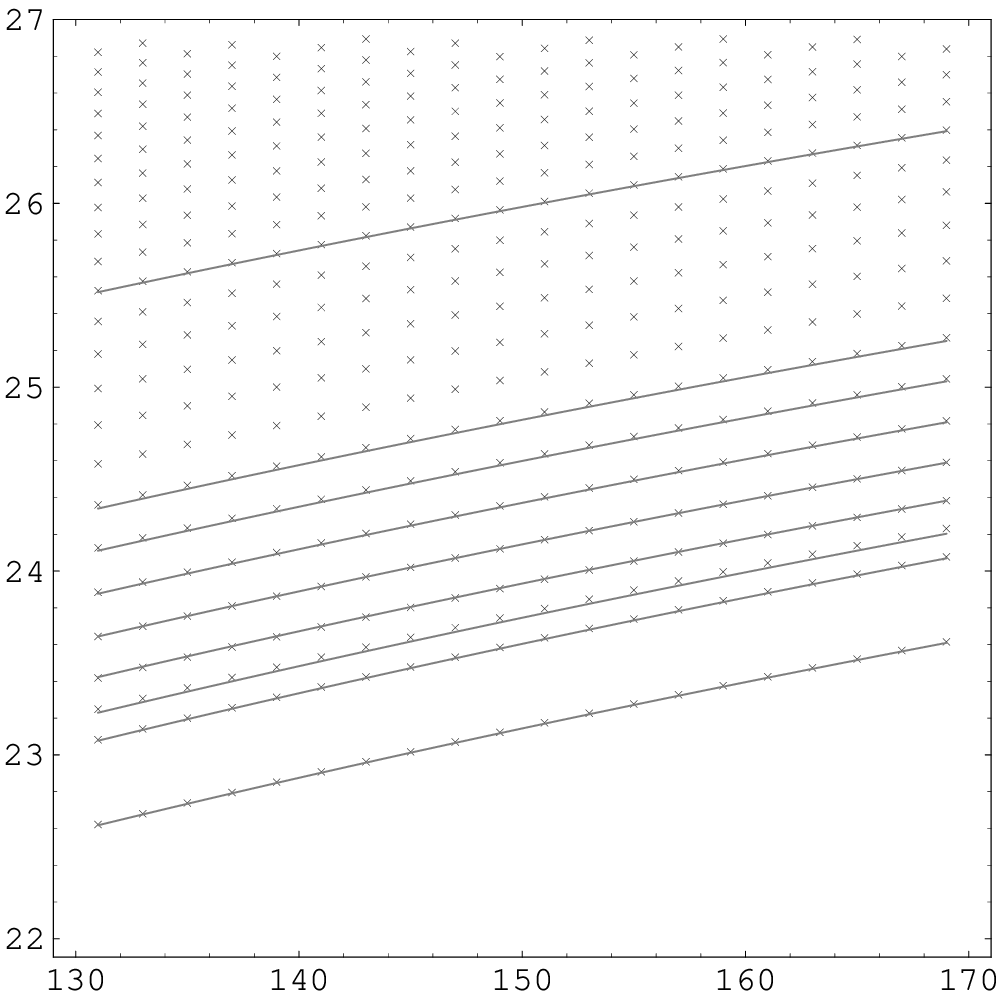}}
\put(46,55){$J$}
\put(-10,100){\rotate{$\cE (J)$}}
\end{picture}
\end{center}
\vspace{-6cm}
\caption{\label{G2energy150} Same as in Fig.\ \ref{G2energy} but for
large $J$. The spectrum is cut at the top where it coincides with the
eigenvalues of the integrable Hamiltonian $\cH_0$.}
\end{figure}

Making use of these results we design the following formulae which
describe well (see Figs.\ \ref{G2energy150}, \ref{G2energy}) the
spectrum of the perturbed Hamiltonian
\begin{eqnarray}
\label{EnerQ2app}
\cE (J, m) &=& \cE_0 (J, 0) - \Delta (m),\quad\mbox{for}\quad m = 0, 1,\\
\cE (J, m) &=& \cE_0 (J, q(m)) - \Delta (m),\quad\mbox{for}\quad m \geq 2 .
\end{eqnarray}
Here $\cE_0 (J, 0)$ is the ground state energy (\ref{Qoenergy})
and $\cE_0 (J, q(m))$ is Eq.\ (\ref{EnergyBottom}) with the
$q (m)$ trajectories deduced from the quantization condition
(\ref{QuantizedQbottom}). The function $\Delta$ is
\begin{eqnarray}
\Delta (0) = 0.54,
\qquad
\Delta (1) = 0.08,
\qquad
\Delta (m \geq 2)
= \left( \delta_0 - \delta (m - 2) \right)
\theta \left( \delta_0 - \delta (m - 2) \right),
\end{eqnarray}
with $\delta_0 = 0.15$ and $\delta = 0.01$. It serves to generate a
shift of the non-perturbed energy trajectories. The top of the
spectrum coincides with the $\cE_0$ trajectories.

\section{Quark-gluon sector.}

Although we have started with the pure gluonic sector, chronologically,
it is the non-singlet quark-gluon-quark sector which has been first
addressed within the context of the transversely polarized structure
function $g_2 (x_{\rm Bj})$. The first study which correctly identified
a complete basis of operators was by Shuryak and Vainshtein and they
have calculated a lowest anomalous dimension \cite{ShuVan82}.
The evolution equation for the correlation function $Y$ in Eq.\
(\ref{gTtoY}) which is defined as the Fourier transform
\begin{equation}
\label{Ydefintion}
Y(x_1, x_3) = \int \frac{d \kappa_1}{2 \pi} \frac{d \kappa_3}{2 \pi}
e^{i x_1 \kappa_1 - i x_3 \kappa_3}
\langle h |
\cY (\kappa_1, 0, \kappa_3)
| h \rangle
\end{equation}
of the $C$-even combination
\begin{equation}
\cY (\kappa_1, 0, \kappa_3)
= \frac{1}{2}
\left\{
{^+\! \cS} (\kappa_1, 0, \kappa_3) +
{^-\! \cS} (- \kappa_3, 0, - \kappa_1)
\right\} ,
\end{equation}
of the three-particle quark-gluon-quark operators
\begin{equation}
{^\pm\! \cS} (\kappa_1, \kappa_2, \kappa_3)
= \frac{1}{2}
\bar{\psi}(\kappa_3 n) i \g \gamma_+
\left[ i \widetilde G^\perp_{\sigma +} (\kappa_2 n)
\pm
\gamma_5 G^\perp_{\sigma +}(\kappa_2 n)
\right]
\psi(\kappa_1 n) ,
\end{equation}
has been derived in Refs.\ \cite{BukKurLip84,BukKurLip83} as well as the
anomalous dimensions of the local operators were calculated. Consequently
this was redone in a number of studies \cite{NonLoc,NonSin} with almost
the same results.

\subsection{Evolution equation.}

In spite of the available results \cite{BukKurLip83} let us discuss
the evolution equation from our present point of view. The function
$Y$ can be decomposed according to this discussion as $Y = {^+\! S}
+ {^-\! S}$. The operators $^{\pm}\! \cS$ are related to each other
by the charge conjugation and, therefore, they evolve autonomously. The
evolution equation for the functions ${^\pm\! S}$ has the same form
as Eq.\ (\ref{EvolutionEquation}). For ${^+\! S}$ the total kernel is
a sum of the ones in subchannels
\begin{eqnarray}
{\mbox{\boldmath$K$}^{^+\! S}}
\!\!\!&=&\!\!\!
{^{qq}\widehat{K}^{V}_{(8)}} (x_1, - x_3 | x'_1, - x'_3)
+ {^{qg}\widehat{K}^{T}_{(3)}} (x_1, x_2 | x'_1, x'_2)
+ {^{qg}\widehat{K}^{V}_{(3)}} (x_2, - x_3 | x'_2, - x'_3)
\nonumber\\
&-&\!\!\! {\scriptstyle\frac{1}{4}}\beta_0
\delta (x_1 - x'_1) \delta (x_3 - x'_3) ,
\end{eqnarray}
and their explicit form can be found in Ref.\ \cite{Bel99a}. Analogous
expression holds for ${^-\! S}$ with $T$ and $V$ subscripts being
interchanged in the quark-gluon kernels. Since these kernels are
equivalent we consider in the following only $^+\! S$. The pair-wise
kernels contain both $C_F$ and $C_F - C_A/2$ colour structures and
this presents one of the sources of complication to find the
analytical solution. Since for QCD $(C_A - 2 C_F)/N_c = 1/9$ is small
we can anticipate that large $N_c$ approximation presents a good
(within $10\%$ accuracy) starting point for solving the three-particle
problem.

Taking the Mellin moments (\ref{Moments}) we can write the evolution
equation in large $N_c$ limit for the moments
\begin{equation}
\frac{d}{d \ln Q^2} {^+\! S}_j^J = - \frac{\alpha_s}{4 \pi}
\sum_{k = 0}^{J}
{\mbox{\boldmath${\mit\Gamma}$}_{jk}^S} (J)
{^+\! S}_k^J ,
\qquad
{\mbox{\boldmath${\mit\Gamma}$}_{jk}^S} (J)
= N_c {\mbox{\boldmath${\gamma}$}_{jk}^S} (J),
\end{equation}
with the large-$N_c$ anomalous dimension matrix \cite{BukKurLip83,NonSin}
\begin{eqnarray}
\label{QGQanomalousDim}
{\mbox{\boldmath${\gamma}$}_{jk}^S}
\!\!\!&=&\!\!\! \delta_{jk}
\left(
\psi (j + 1) + \psi (j + 4)
+ \psi (J - j + 2) + \psi (J - j + 3)
- 4 \psi (1) - \frac{3}{2}
\right) \nonumber\\
&-&\!\!\! \theta_{j, k + 1 }
\frac{( k + 2 )_2}{( j - k )( j + 2 )_2}
- \theta_{k, j + 1}
\frac{( J - k + 1 )_2}{( k - j )( J - j + 1 )_2} .
\end{eqnarray}

For large $N_c$ we have ${^{qq}\widehat{K}^{V}_{(8)}} \propto 1/N_c$
and the eigenvalues of the other kernels read, up to terms suppressed
in $N_c$,
\begin{eqnarray}
\label{EigenvaluesKqg}
&&\!\!\!\!\!\!\!\!\!\!\!\!\!\!\!\!\int dx_1 dx_2 P^{\left( 1, 2 \right)}_j
\left( \frac{x_1 - x_2}{x_1 + x_2} \right) \,
\left\{
{
{^{qg}\widehat{K}^{V}_{(3)}}
\atop
{^{qg}\widehat{K}^{T}_{(3)}}
}
\right\}
\left( x_1, x_2 | x'_1, x'_2 \right) \nonumber\\
&&\qquad\qquad\qquad\qquad\qquad
= \frac{N_c}{2}
\left\{
{ \psi (j + 1) + \psi (j + 4) - 2 \psi (1) - \frac{5}{3}
\atop
\psi (j + 2) + \psi (j + 3) - 2 \psi (1) - \frac{5}{3} }
\right\}
P^{\left( 1, 2 \right)}_j
\left( \frac{x'_1 - x'_2}{x'_1 + x'_2} \right) .
\end{eqnarray}
Therefore, similarly to the previous study we can replace the original 
problem of the diagonalization of the three-particle kernel
${\mbox{\boldmath$K$}^{^+\! S}} \equiv \frac{N_c}{2} {\mbox{\boldmath$H$}}
\to \frac{N_c}{2} \cH$ by the solution of the Schr\"odinger equation
\begin{equation}
\label{ScroedinerEq}
{\cH} {\mit\Psi} = \cE {\mit\Psi},
\end{equation}
with the Hamiltonian
\begin{equation}
{\cH} = h_{12} + h_{23} - \frac{3}{2},
\end{equation}
where the pair-wise Hamiltonians
\begin{equation}
\label{QCDTwoSiteHamilton}
h_{12} = \psi \left( {\hat J}_{12} + \frac{3}{2} \right)
+ \psi \left( {\hat J}_{12} - \frac{3}{2} \right) - 2 \psi (1),
\qquad
h_{23} = \psi \left( {\hat J}_{23} + \frac{1}{2} \right)
+ \psi \left( {\hat J}_{23} - \frac{1}{2} \right) - 2 \psi (1) ,
\end{equation}
can be easily read from Eq.\ (\ref{EigenvaluesKqg}).

\subsection{Inhomogeneous open spin chain.}

It turns out that the problem (\ref{ScroedinerEq}-\ref{QCDTwoSiteHamilton})
is exactly solvable\footnote{See Ref.\ \cite{KarKir99} for spin models
appeared in the solution of the quark-gluon reggeon interaction.}.
Let us consider the inhomogeneous spin chain which is characterized
by `inhomogeneities' $\delta_\ell$. The monodromy matrix is defined
as before but with the shifted spectral parameters
\begin{equation}
T_b ( \lambda ) = R_{a_1,b} ( \lambda - \delta_1 )
R_{a_2,b} ( \lambda - \delta_2 ) R_{a_3,b} ( \lambda - \delta_3 ) .
\end{equation}
The generating function of the conserved integrals of motion, i.e.\
the transfer matrix, reads
\cite{Skl88}
\begin{equation}
\label{OpenTransfer}
t_b ( \lambda )
= {\rm tr}_b\, K^+ ( \lambda ) T_b ( \lambda )
K^- ( \lambda ) T_b^{- 1} ( - \lambda ),
\end{equation}
with boundary reflection matrices $K^\pm$ which we set equal to unit
matrix in order to ensure the conformal invariance of $t_b$. The matrices 
(\ref{OpenTransfer}) obey the commutation relation (\ref{Commutativity}).
The shifted and renormalized auxiliary transfer matrix
$\prod_{\ell = 1}^{3} \Big\{ (\lambda - \delta_\ell)
(\lambda - \delta_\ell  -1) - {\mbox{\boldmath$\hat \jmath$}}^2_\ell
\Big\} t_{\frac{1}{2}} (\lambda - {\scriptstyle\frac{1}{2}})
\to t_{\frac{1}{2}} (\lambda)$ \cite{Skl88}
\begin{equation}
t_{\frac{1}{2}} ( \lambda )
= {\rm tr}_{\frac{1}{2}}\,
L_{a_1} ( \lambda - \delta_1 ) L_{a_2} ( \lambda - \delta_2 )
L_{a_3} ( \lambda - \delta_3 )
\sigma_2
L_{a_3}^{\sf t} ( - \lambda - \delta_3 )
L_{a_2}^{\sf t} ( - \lambda - \delta_2 )
L_{a_1}^{\sf t} ( - \lambda - \delta_1 )
\sigma_2 ,
\end{equation}
which is an even function in $\lambda$, has the following expansion in
the rapidity\footnote{We would like to thank G. Korchemsky for a discussion
on this point (see also recent Ref.\ \cite{DerKorMan99}).}
\begin{equation}
t_{\frac{1}{2}} ( \lambda )
= {\mit\Omega} ( \lambda )
- (4 \lambda^2 - 1)
\left( \lambda^2 - \delta_2^2 - {\mbox{\boldmath$\hat \jmath$}}^2_2 \right)
{\mbox{\boldmath$\hat J$}}^2
- \frac{1}{2} (4 \lambda^2 - 1) \cQ_\delta ,
\end{equation}
where ${\mit\Omega} ( \lambda )$ is the c-number function
\begin{eqnarray*}
{\mit\Omega} ( \lambda )
= - 2 \prod_{\ell = 1}^{3}
\left(
\lambda^2 + {\mbox{\boldmath$\hat \jmath$}}^2_\ell - \delta^2_\ell
\right)
+ (4 \lambda^2 - 1)
\sum_{\ell = 1}^{3} {\mbox{\boldmath$\hat \jmath$}}^2_\ell
\left( \lambda^2 + \delta^2_1 + \delta^2_3 - \delta^2_\ell \right)
+ (4 \lambda^2 - 1)
{\mbox{\boldmath$\hat \jmath$}}^2_1 {\mbox{\boldmath$\hat \jmath$}}^2_3 ,
\end{eqnarray*}
${\mbox{\boldmath$\hat J$}}^2$ is the total Casimir operator
and $\cQ_\delta$ is the non-trivial integral of motion
\begin{equation}
\label{DeltaCharge}
\cQ_\delta
= [ {\mbox{\boldmath$\hat J$}}^2_{12},
{\mbox{\boldmath$\hat J$}}^2_{23} ]_+
- 2 ( \delta^2_1 - \delta^2_2 )
{\mbox{\boldmath$\hat J$}}^2_{23}
- 2 ( \delta^2_3 - \delta^2_2 )
{\mbox{\boldmath$\hat J$}}^2_{12}
+ 8 i \delta_2 \epsilon_{ijk} \hat J^i_1 \hat J^j_2 \hat J^k_3 .
\end{equation}
As will be clear later on, only $\delta_2 = 0$ spin chains appear in
QCD within the twist-three context. Therefore, we set in what follows
$\delta_2 = 0$.

The Hamiltonian which commutes with the charges found above is deduced
from the fundamental transfer matrix and reads \cite{Skl88} $\cH_\delta
= h_{12} + h_{23}$, with the two-site Hamiltonians expressed in terms of 
the fundamental $R$-matrices $h_{a_1,a_2} = R_{a_1,a_2} (- \delta)
R'_{a_1,a_2} (- \delta)$ so that explicitly
\begin{equation}
h_{12} = \psi \left( {\hat J}_{12} + \delta_1 \right)
+ \psi \left( {\hat J}_{12} - \delta_1 \right) - 2 \psi (1),
\qquad
h_{23} = \psi \left( {\hat J}_{23} + \delta_3 \right)
+ \psi \left( {\hat J}_{23} - \delta_3 \right) - 2 \psi (1),
\end{equation}
provided we set in the bundle $R_{ab} (\lambda)$ (\ref{YBbundle}) the
function $f (\nu, \lambda)$ as was done after Eq.\ (\ref{TwoSiteHamilton}).
The commutation rule for the $h_{a_1, a_2}$ with $\cQ_\delta$ is
$[\cQ_\delta, h_{12}]_- = 2 [ {\mbox{\boldmath$\hat J$}}^2_{23} ,
{\mbox{\boldmath$\hat J$}}^2_{12} ]_-$ and $[\cQ_\delta, h_{23}]_-
= 2 [ {\mbox{\boldmath$\hat J$}}^2_{12} ,
{\mbox{\boldmath$\hat J$}}^2_{23} ]_-$. So that the property
$[\cQ_\delta, \cH]_- = 0$ is obvious.

Setting $\delta_1 = 3/2$, $\delta_3 = 1/2$ we get the QCD result
(\ref{QCDTwoSiteHamilton}) and $\cQ_S$ charge of Ref.\ \cite{BraDerMan98}.
The conformal weights of the end quarks are $\nu_1 = \nu_3 = 1$ and
$\nu_2 = 2$ for the gluon.

\subsection{Master equation and integral of motion.}

As before we can solve now the simplified problem
\begin{equation}
\label{EqQS}
\cQ_S {\mit\Psi} = q_S {\mit\Psi},
\end{equation}
instead of Eq.\ (\ref{ScroedinerEq}), where $\cQ_S$ is Eq.\
(\ref{DeltaCharge}) with $\delta$'s set earlier.
The main steps of the solution of this problem are the same as we
have pursued in the preceding section and in Ref.\ \cite{Bel99b} so
that we just outline them briefly.

First it is convenient to take the polynomials, which diagonalize
simultaneously ${\mbox{\boldmath$\hat J$}}^2$ and
${\mbox{\boldmath$\hat J$}}^2_{12}$,
\begin{equation}
\cP_{J; j}
(\theta_1, \theta_2 | \theta_3)
= \frac{\varrho_j^{- 1}}{\sqrt{2}}
\frac{(j + 2)}{(j + 1)(j + 3)}
\frac{\Gamma (j + 3)}{\Gamma (2j + 4)}
\frac{\Gamma (J + j + 5)}{\Gamma^{1/2} (2 J + 6)}
\theta_{12}^J \, \theta^{j - J}
{_2F_1} \left( \left.
{ j - J , j + 2 \atop 2j + 5 }
\right| \theta \right) ,
\end{equation}
with $\theta \equiv \frac{\theta_{12}}{\theta_{32}}$, as basis vectors
since the off-diagonal elements of the $\cQ_S$ in this basis will be
the same as for $\cQ_T$ in \cite{Bel99b} so that only diagonal elements
will be slightly modified. Expanding the eigenfunctions ${\mit\Psi}$
in the way it was done in (\ref{EigenFunExpansion})
\begin{equation}
\label{RC1}
{\mit\Psi} = \sum_{j = 0}^{J} {\mit\Psi}_j
\cP_{J; j} (\theta_1, \theta_2 | \theta_3),
\end{equation}
it is easy to show that Eq.\ (\ref{EqQS}) is equivalent to the three-term
recursion relation
\begin{equation}
\label{MasterEq}
(2j + 3) {\mit\Upsilon}_{j + 1}
+
(2j + 5) {\mit\Upsilon}_{j - 1}
+
\varrho_j^2\, (2j + 3)(2j + 5)
\left(
{[ \cQ_S ({\scriptstyle\frac{3}{2}}, {\scriptstyle\frac{1}{2}})]}_{j,j}
- q_S
\right) {\mit\Upsilon}_j
= 0 ,
\end{equation}
with boundary conditions ${\mit\Upsilon}_{- 1} = {\mit\Upsilon}_{J + 1}
= 0$, and where
\begin{eqnarray}
[\cQ_S (\delta_1, \delta_3)]_{j,j} \!\!\!&=&\!\!\! - 2 \delta_3^2
\left( j + \frac{3}{2} \right) \left( j + \frac{5}{2} \right)
+ \left\{
\left( j + \frac{3}{2} \right) \left( j + \frac{5}{2} \right)
- \delta_1^2
\right\} \nonumber\\
&\times&\!\!\!
\left\{
\frac{3}{4}
- \left( j + \frac{3}{2} \right) \left( j + \frac{5}{2} \right)
+ \left( J + \frac{5}{2} \right) \left( J + \frac{7}{2} \right)
+ \frac{3}{4}
\frac{ \left( J + \frac{5}{2} \right) \left( J + \frac{7}{2} \right) }
{\left( j + \frac{3}{2} \right) \left( j + \frac{5}{2} \right) }
\right\} .
\end{eqnarray}
Here the new expansion coefficients are introduced as follows
\begin{equation}
{\mit\Psi}_j \equiv \varrho_j {\mit\Upsilon}_j ,
\qquad\mbox{with}\qquad
\varrho_j =
\left[
\frac{(j + 1)^3 (j + 3)^3}{(j + 2)^3} (J - j + 1) (J + j + 5)
\right]^{- 1/2} .
\end{equation}

However, due to the loss of any symmetry of the integral of motion
w.r.t.\ the permutation of the sites it is instructive to study
the recurrence relation based on the expansion of the eigenfunctions
in the basis $\cP_{J; j} (\theta_3, \theta_2 | \theta_1)$
\begin{equation}
\label{RC2}
{\mit\Psi} = \sum_{j = 0}^{J} {\mit\Phi}_j
\cP_{J; j} (\theta_3, \theta_2 | \theta_1),
\end{equation}
where
\begin{equation}
\cP_{J; j} (\theta_3, \theta_2 | \theta_1)
= \sqrt{2} \, \varsigma_j^{- 1}
\frac{\Gamma (j + 3)}{\Gamma (2j + 5)}
\frac{\Gamma (J + j + 5)}{\Gamma^{1/2} (2 J + 6)}
\theta_{32}^J \, \theta^{j - J}
{_2F_1} \left( \left.
{ j - J , j + 2 \atop 2j + 5 }
\right| \theta \right) ,
\quad\mbox{with}\quad
\theta \equiv \frac{\theta_{32}}{\theta_{12}} .
\end{equation}
Introducing
\begin{equation}
{\mit\Phi}_j \equiv \varsigma_j {\mit\Xi}_j,
\quad\mbox{with}\quad
\varsigma_j =
\left[ (j + 1) (j + 2) (j + 3) (J - j + 1) (J + j + 5) \right]^{- 1/2} ,
\end{equation}
we have the recursion relation for ${\mit\Xi}_j$
\begin{equation}
(2j + 3) {\mit\Xi}_{j + 1}
+
(2j + 5) {\mit\Xi}_{j - 1}
+
\varsigma_j^2\, (2j + 3)(2j + 5)
\left(
{[ \cQ_S ({\scriptstyle\frac{1}{2}}, {\scriptstyle\frac{3}{2}})]}_{j,j}
- q_S
\right) {\mit\Xi}_j
= 0 .
\end{equation}

From the condition of existence of a solution to the recursion
relations which satisfies the boundary conditions we find that the
allowed values of $q_S$ asymptotically lie in the band $0 \leq
q_S/J^4 \leq \frac{1}{2}$. The upper boundary is achieved at
$j_{\rm max} = \frac{1}{\sqrt{2}} J$. In complete analogy with
$\cQ_T$ \cite{Bel99b}, the spectrum of $\cQ_S$ can be described by
means of two different sets of trajectories. The first ones which
behave as $J^2$ and another ones as $J^4$, at large $J$.

Let us turn to the first possibility. One can immediately find from
Eq.\ (\ref{MasterEq}) the exact lowest trajectory for the charge
\begin{equation}
\label{qSexact}
q_S^{\rm exact} (J)
= \left( J + \frac{5}{2} \right) \left( J + \frac{7}{2} \right)
+ \frac{5}{8}.
\end{equation}
For other levels we construct an effective WKB approximation. For the case 
at hand the classical motion is allowed for the whole interval of $j$ 
except of the vicinities of the reflection points $j_{\rm end} \sim 1, J$,
which will be parametrized by the continuous parameter $\tau \equiv j/J$.
Introducing the new function as ${\mit\Upsilon}_j = J (- 1)^j \upsilon
(\tau)$ and $q_S^\star = q_S/J^2$, we have from (\ref{MasterEq}) the
differential equation
\begin{equation}
\tau^2 (1 - \tau^2) \upsilon'' (\tau) - \tau (1 - \tau^2) \upsilon' (\tau)
+ 2 (q_S^\star - 1) \upsilon (\tau) = 0 ,
\end{equation}
the solution to which reads
\begin{equation}
\label{QGQwkbSol1}
\upsilon (\tau) = C^{(+)} \tau^{1 + 2 i \eta_S}
{_2F_1} \left( \left.
{ \frac{1}{2} + i \eta_S , - \frac{1}{2} + i \eta_S \atop 1 + 2 i \eta_S }
\right| \tau^2 \right)
+ C^{(-)} \tau^{1 + 2 i \eta_S}
{_2F_1} \left( \left.
{ \frac{1}{2} - i \eta_S , - \frac{1}{2} - i \eta_S \atop 1 - 2 i \eta_S }
\right| \tau^2 \right) ,
\end{equation}
where $\eta_S = \frac{1}{2} \sqrt{2 q_S^\star - 3}$. Analogously for
${\mit\Xi}_j = J (- 1)^j \xi (\tau)$ we have the equation
\begin{equation}
\tau^2 (1 - \tau^2) \xi'' (\tau) - \tau (1 - \tau^2) \xi' (\tau)
+ 2 ( q_S^\star - 1 + 4 \tau^2 ) \xi (\tau) = 0 ,
\end{equation}
with the solution
\begin{equation}
\label{QGQwkbSol2}
\xi (\tau) = \widetilde C^{(+)} \tau^{1 + 2 i \eta_S}
{_2F_1} \left( \left.
{ \frac{3}{2} + i \eta_S , - \frac{3}{2} + i \eta_S \atop 1 + 2 i \eta_S }
\right| \tau^2 \right)
+ \widetilde C^{(-)} \tau^{1 + 2 i \eta_S}
{_2F_1} \left( \left.
{ \frac{3}{2} - i \eta_S , - \frac{3}{2} - i \eta_S \atop 1 - 2 i \eta_S }
\right| \tau^2 \right) .
\end{equation}
Here the arbitrary complex constants $C^{(\pm)}$ and
$\widetilde C^{(\pm)}$ have to be fixed from a sewing procedure with
the solutions ${\mit\Psi}_j$ and ${\mit\Phi}_j$ close to the
boundaries.

For $J \gg J - j \sim 1$ we can easily solve (\ref{MasterEq})
\begin{equation}
\label{QGQBoundaryOne}
{\mit\Upsilon}_j = {\mit\Xi}_j = (- 1)^j (J - j + 1) .
\end{equation}
To solve the recursion relation for $J \gg j \sim 1$ is more difficult
but we accept the same strategy as for Eq.\ (\ref{LimitRecOne}). However,
since the permutation symmetry is lost we have to modify the procedure
accordingly. Namely, the expansion coefficients in Eqs.\
(\ref{RC1},\ref{RC2}) are related to each other by
\begin{equation}
{\mit\Psi}_j = \sum_{k = 0}^{J} W_{kj} {\mit\Phi}_k,
\qquad
{\mit\Phi}_j = \sum_{k = 0}^{J} W_{jk} {\mit\Psi}_k,
\end{equation}
with the Racah coefficients $W_{jk}$, $\cP_{J; j} ( \theta_3, \theta_2 |
\theta_1 ) = \sum_{k = 0}^{J} W_{jk} (J) \cP_{J; k} ( \theta_1, \theta_2
| \theta_3 )$, which can be read off from the general result of Ref.\
\cite{Bel99b}. Then, for the region in question the solutions can be
written as
\begin{equation}
\left\{
{ {\mit\Upsilon}_j \atop {\mit\Xi}_j }
\right\}
= (- 1)^J (j + 3) \int\limits_{0}^{1} d \tau
P^{(1,2)}_j (2 \tau - 1)
\left\{
\begin{array}{r}
\frac{(j + 1)(j + 3)}{j + 2} \xi (\tau) \\
(j + 2) \upsilon (\tau)
\end{array}
\right\} .
\end{equation}
Evaluating the integrals we obtain (see Ref.\ \cite{Bel99b})
\begin{eqnarray}
\label{QGQBoundaryTwo1}
{\mit\Upsilon}_j = (- 1)^J \frac{(j + 1)(j + 3)}{j + 2}
&&\!\!\!\!\!\!\!\!\!
\Bigg\{
\widetilde C^{(+)}
\Bigg[
{_2F_1} \left( \left.
{ \frac{3}{2} + i \eta_S , - \frac{3}{2} + i \eta_S \atop 1 + 2 i \eta_S }
\right| 1 \right)
+
(- 1)^j
\frac{\Gamma \left( \frac{3}{2} + i \eta_S \right)
\Gamma \left( \frac{5}{2} + j - i \eta_S \right)
}{\Gamma \left( \frac{3}{2} - i \eta_S \right)
\Gamma \left( \frac{5}{2} + j + i \eta_S \right)} \nonumber\\
\times&&\!\!\!\!\!\!\!\!\!\! {_4F_3} \left( \left.
{ - \frac{3}{2} + i \eta_S ,\ - \frac{1}{2} + i \eta_S ,\
\frac{3}{2} + i \eta_S ,\ \frac{3}{2} + i \eta_S
\atop
- \frac{3}{2} - j + i \eta_S ,\
\frac{5}{2} + j + i \eta_S ,\
1 + 2 i \eta_S } \right| 1 \right)
\Bigg]
+
\widetilde C^{(-)}
[ \eta_S \to - \eta_S]
\Bigg\}, \nonumber\\
\end{eqnarray}
and
\begin{eqnarray}
\label{QGQBoundaryTwo2}
{\mit\Xi}_j = (- 1)^J (j + 2)
&&\!\!\!\!\!\!\!\!\!
\Bigg\{
C^{(+)}
\Bigg[
{_2F_1} \left( \left.
{ \frac{1}{2} + i \eta_S , - \frac{1}{2} + i \eta_S \atop 1 + 2 i \eta_S }
\right| 1 \right)
+
(- 1)^j
\frac{\Gamma \left( \frac{3}{2} + i \eta_S \right)
\Gamma \left( \frac{5}{2} + j - i \eta_S \right)
}{\Gamma \left( \frac{3}{2} - i \eta_S \right)
\Gamma \left( \frac{5}{2} + j + i \eta_S \right)} \nonumber\\
\times&&\!\!\!\!\!\!\!\!\!\! {_4F_3} \left( \left.
{ - \frac{1}{2} + i \eta_S ,\ - \frac{1}{2} + i \eta_S ,\
\frac{1}{2} + i \eta_S ,\ \frac{3}{2} + i \eta_S
\atop
- \frac{3}{2} - j + i \eta_S ,\
\frac{5}{2} + j + i \eta_S ,\
1 + 2 i \eta_S } \right| 1 \right)
\Bigg]
+
C^{(-)}
[ \eta_S \to - \eta_S]
\Bigg\}, \nonumber\\
\end{eqnarray}

Comparing Eqs.\ (\ref{QGQwkbSol1}) and
(\ref{QGQBoundaryOne},\ref{QGQBoundaryTwo1},\ref{QGQBoundaryTwo2}) in
the region of their overlap we find
\begin{equation}
\frac{C^{(+)}}{C^{(-)}}
= - \frac{
{_2F_1} \left( \left.
{ \frac{1}{2} - i \eta_S , - \frac{1}{2} - i \eta_S \atop 1 - 2 i \eta_S }
\right| 1 \right)}{
{_2F_1} \left( \left.
{ \frac{1}{2} + i \eta_S , - \frac{1}{2} + i \eta_S \atop 1 + 2 i \eta_S }
\right| 1 \right) } ,
\qquad\qquad
\frac{\widetilde C^{(+)}}{\widetilde C^{(-)}}
= - \frac{
{_2F_1} \left( \left.
{ \frac{3}{2} - i \eta_S , - \frac{3}{2} - i \eta_S \atop 1 - 2 i \eta_S }
\right| 1 \right)}{
{_2F_1} \left( \left.
{ \frac{3}{2} + i \eta_S , - \frac{3}{2} + i \eta_S \atop 1 + 2 i \eta_S }
\right| 1 \right) } ,
\end{equation}
and the quantization condition for the eigenvalues $q_S$ (cf.\
\cite{BraDerMan98})
\begin{equation}
4 \eta_S \ln J - \arg
\frac{\widetilde C^{(-)}}{\widetilde C^{(+)}}
\frac{C^{(+)}}{C^{(-)}}
\frac{\Gamma^2 \left( \frac{3}{2} + i \eta_S \right)}{
\Gamma^2 \left( \frac{3}{2} - i \eta_S \right)}
= 2 \pi m ,
\end{equation}
where $m \in \bfZ_+$.

For the upper part of the spectrum $q_S \propto J^4$, the classical motion
is concentrated in the vicinity of the point $j_{\rm max}$. Looking the
solution to the recursion relation in the form of the series (\ref{WKBexp})
we find immediately that the differential equations resulted from these
are almost the same, to the required accuracy, as for the chiral-odd sector
so that we can immediately adapt the results of Ref.\ \cite{Bel99b}.
Therefore, the WKB expansion of the charge reads
\begin{equation}
\label{WKBexpCharge}
q_S (J) = \sum_{\ell = 0}^{\infty} \frac{q_S^{(\ell)}}{J^\ell},
\end{equation}
with the expansion coefficients
\begin{eqnarray}
q_S^{(0)} (n)
&=& \frac{1}{2} , \nonumber\\
q_S^{(1)} (n)
&=& 6 - \frac{1}{\sqrt{2}} - \sqrt{2} n , \nonumber\\
q_S^{(2)} (n)
&=& \frac{1}{16} \left( 403 - 72 \sqrt{2} \right)
+ \frac{1}{8} \left( 11 - 72 \sqrt{2} \right) n
+ \frac{11}{8} n^2 ,\\
\dots ,\quad\ \ && \nonumber
\end{eqnarray}
with $n =0,1,\dots$. It differs form the $q_T$ expression in Ref.\
\cite{Bel99b} only at the level of $\cO (J^{- 2})$ corrections.

\subsection{Energy spectrum.}

\begin{figure}[p]
\unitlength1mm
\begin{center}
\vspace{-1cm}
\hspace{0cm}
\begin{picture}(100,155)(0,0)
\put(0,60){\insertfig{9}{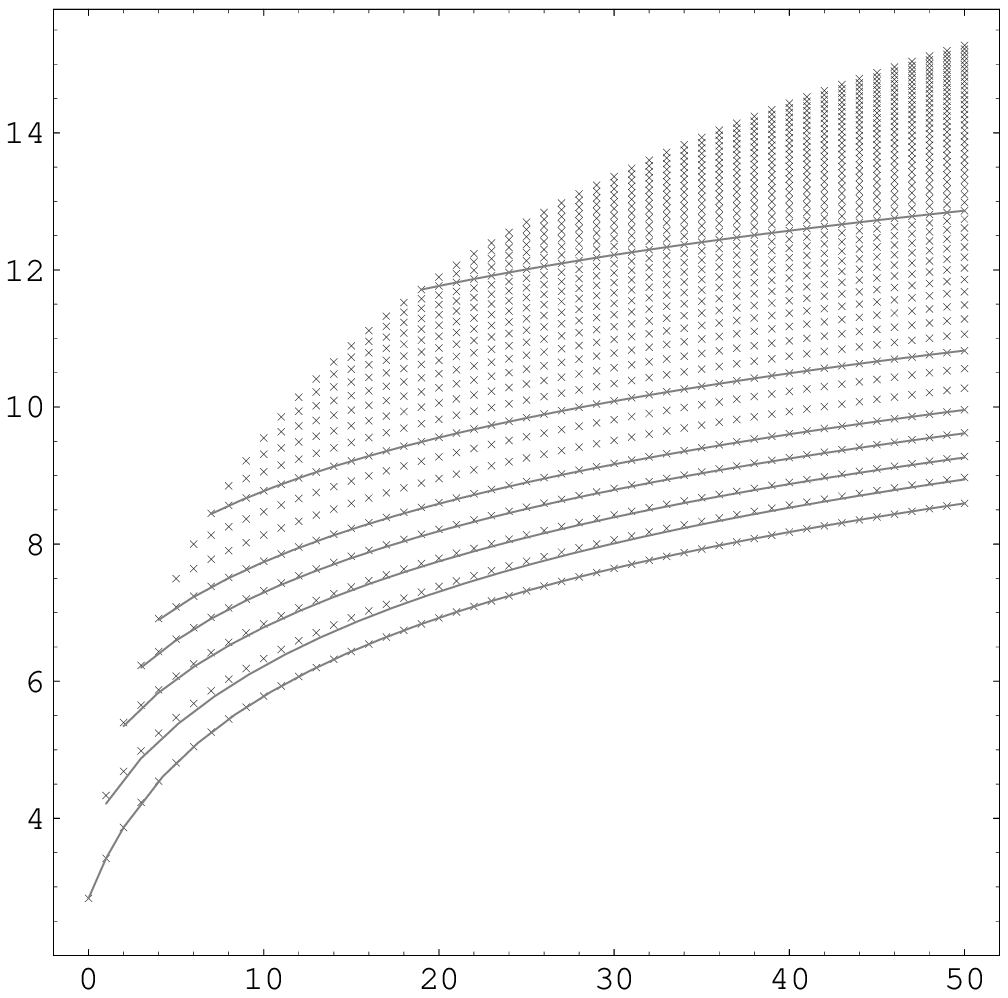}}
\put(46,55){$J$}
\put(-10,100){\rotate{$\cE (J)$}}
\end{picture}
\end{center}
\vspace{-6cm}
\caption{\label{QGQenergy}
The analytical trajectories from the Eqs.\
(\ref{LowestEnergy},\ref{TopAfterGap}) and the numerically
diagonalized anomalous dimension matrix (\ref{QGQanomalousDim}).}
\end{figure}

\begin{figure}[p]
\unitlength1mm
\begin{center}
\vspace{-1cm}
\hspace{0cm}
\begin{picture}(100,155)(0,0)
\put(0,60){\insertfig{9}{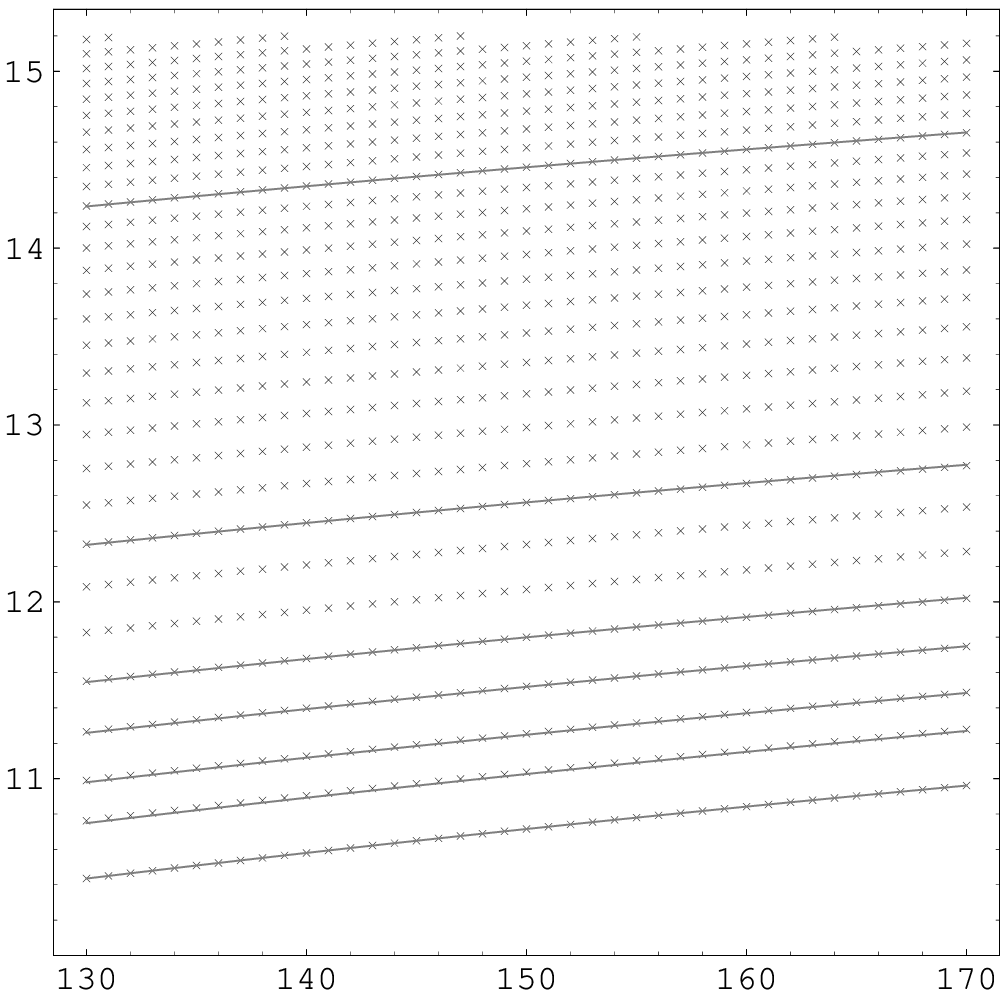}}
\put(46,55){$J$}
\put(-10,100){\rotate{$\cE (J)$}}
\end{picture}
\end{center}
\vspace{-6cm}
\caption{\label{QGQenergy150} Same as in Fig.\ \ref{QGQenergy} but
for large $J$. The spectrum is cut from the top where it is almost
continuous.}
\end{figure}

The knowledge of eigenfunctions allows to find the energy of the
quark-gluon-quark system. For the trajectories starting from the top
of the spectrum we have
\begin{equation}
\cE (J, q_S) = \ln q_S/2 - 4 \psi (1) - \frac{3}{2} + \cO (J^{- 1}) ,
\end{equation}
with $q_S$ taken from Eq.\ (\ref{WKBexpCharge}).

For the levels behaving as $\cE \propto 2 \ln J$ we can find the
exact lowest trajectory \cite{AliBraHil91} corresponding to the
charge (\ref{qSexact})
\begin{equation}
\label{LowestEnergy}
\cE (J) = \psi (J + 3) + \psi (J + 4) - 2 \psi (1) - \frac{1}{2} .
\end{equation}
The remainder of the spectrum is described by the formula
\cite{BraDerMan98,Bel99b}
\begin{equation}
\label{TopAfterGap}
\cE (J, q_S)
= 2 \ln J - 4 \psi (1)
+ 2 \, {\rm Re} \, \psi \left( \frac{3}{2} + i \eta_S \right)
- \frac{3}{2} ,
\end{equation}
and it compares quite well with the explicit numerical diagonalization
of Eq.\ (\ref{QGQanomalousDim}) as shown in Figs.\ \ref{QGQenergy}, 
\ref{QGQenergy150}.

\section{Conclusion.}

In this paper we have demonstrated that the one-loop QCD evolution
equations for the three-gluon and quark-gluon-quark correlation
functions can be reduced to (almost) integrable one dimensional lattice
models. The appearance of the integrable interaction deduced from the
Yang-Baxter bundle (\ref{YBbundle}), within the present context, is a
consequence of the logarithmic behaviour of QCD in collinear regime
since the Mellin transform w.r.t.\ momentum of the emitted gluon from
a particle diverges as $\ln J$ for large $J$, i.e.\ for the kinematics
with negligible momentum transfer to a gluon. For finite $J$ it stems 
from the $\psi (J)$-function, i.e.\ the one which appears in local 
Hamiltonians. It is well known that in the region of momenta mentioned 
above, QCD can be replaces by the theory of eikonalized fields which 
should possesses, hence, more symmetries than the original QCD Lagrangian.

Integrability allows to disentangle the energy spectra of the
three-particle systems and an additional hidden charge serves to
count the trajectories giving rise to different scale dependence
of the corresponding multiplicatively renormalizable components of
the correlators. The explicit analytical solutions have been found
in the form of WKB type expansions w.r.t.\ the total conformal spin
$J$ of the system. A few first terms in the series gave a good
accuracy as compared to the numerical results obtained by a brute
force diagonalization of the anomalous dimension matrices. Moreover,
our analytical expressions are valid up to a very low $J$ and, therefore, 
this allows to apply the solutions for realistic calculations of the 
effects of scaling violations in physical observables.

\vspace{1cm}

We would like to thank V.M. Braun and D. M\"uller for a conversation at
the preliminary stages of the work and especially G.P. Korchemsky for a
number of discussions during the visit of the present author at LPTHE
(Orsay) in May 1998. We acknowledge the support from the Alexander von
Humboldt Foundation.

\end{document}